%% file: intermediate_decoding.tex
\documentclass[twocol]{ametsocV6.1}
\usepackage[T1]{fontenc}

\usepackage{tikz}
\usepackage{placeins}
\usepackage{booktabs}
\usetikzlibrary{arrows.meta, positioning, calc}

\usepackage{svg}
\usepackage{import}
\usepackage{stfloats}

\title{Interpretability of AI weather models via intermediate decoding}

\authors{Matthias Beylich,\correspondingauthor{Matthias Beylich, Matthias.Beylich@lmu.de}
Kirsten I. Tempest,
Tobias Selz\thanks{Tobias Selz current affiliation: Karlsruhe Institute of Technology},
George C. Craig
}
\affiliation{Meteorological Institute, LMU Munich}

\abstract{AI weather prediction models are often regarded as black boxes. Despite their impressive skill, there is little understanding of how the input data is processed during inference to produce the forecast. This limitation arises from the use of high-dimensional latent representations in which individual dimensions are no longer directly interpretable in physical terms. To gain insight into how forecasts are constructed, we propose the method of \emph{intermediate decoding} for AI weather models and apply it to GraphCast, Aurora and ArchesWeather. In this approach, latent states from intermediate depths are decoded back into meteorological fields. To do this, we use the decoder that has been trained to output the forecast at the end of processing. Since latent states at different depths do not share a consistent basis, we train affine transformations that align intermediate latent states with the decoder’s expected input space. Applying this framework to the three models yields the following results. First, the skill of forecasts at intermediate stages of processing improves gradually through the processor layers. Second, bypassing all processor blocks shows that the encoder–decoder pair is not an identity mapping returning the input, but instead plays a significant role in the forecast construction. Third, by examining increments of individual processor blocks across model depth, we show that all three models progressively refine the forecast from coarse to finer spatial scales. Finally, we assess the physical consistency of the intermediate decoding forecasts and find that large scales remain physically balanced throughout processing, while small scales become more balanced with each additional processor block.}

\begin{document}

\maketitle

%
%
%
%
%
%

%

\section{Introduction}

Recent years have seen the emergence of data-driven AI weather models that are competitive with, and often surpass, traditional numerical weather prediction models on standard benchmarks. These models still show limitations, however, such as unphysical behavior and overly smooth forecasts \citep{chattopadhyay2023challenges, Selz2026, bonavita_limitations_2024}, but it is often unclear how these limitations arise and how they could be corrected. 

Despite their strong forecast skill, AI weather models remain difficult to interpret. The model architectures, for example, transformers and graph neural networks, are not motivated by physical laws, and the internal state is encoded in a high-dimensional latent space vector that bears no obvious correspondence to observable physical quantities. As a result, such models are often regarded as black boxes \citep{mcgovern2019interpretable}.

This lack of interpretability limits trust in the forecasts and hinders model development. The opacity of the latent space makes it difficult to gain insight into what the models have actually learned. In a pessimistic view, AI weather models merely interpolate statistically among situations seen in the training data. Alternatively, they may implicitly solve the known physical equations. More intriguingly, they might encode a new representation of physics \citep{craig2026physicsaiweathermodels} that, once understood, could provide new insights into atmospheric processes.

Research into the interpretability of AI weather models has broadly followed two main approaches \citep{yang_interpretable_2024}. The first focuses on inherently interpretable, simplified models, while the second applies post hoc techniques to more complex models, including perturbation-based, game-theoretic, and gradient-based methods. But simplified models lack complexity and realism, while post hoc approaches primarily assess the importance of input variables or regions for a model’s predictions. Consequently, most work does not directly investigate the latent space or attempt to understand models at the level of individual neurons, as is common in mechanistic interpretability studies of large language models \citep{saphra-wiegreffe-2024-mechanistic}. Recent efforts have begun to move in this direction, for example by identifying latent features in GraphCast using sparse autoencoders \citep{macmillan_towards_2025}. However, these approaches are only a first step toward a holistic understanding of how predictions are formed. 

Modern AI weather models are based on deep neural networks, and a key question is how the representation evolves through the stages of processing. The processor typically employs a residual network architecture \citep{He:2015tt}, in which a residual state is propagated through the depth of the model while individual processing blocks incrementally update this state. 

Different paradigms have been proposed for how the state is modified by the sequence of processor blocks. The ensemble view interprets deep networks as a collection of many short computational paths that jointly contribute to the final prediction \citep{veit2016residual}. Large changes to the representation can occur at any step in the processing. On the other hand, the iterative refinement view emphasizes how representations evolve within the residual stream \citep{jastrzebski_residual_2018, greff2017highway}. In this perspective, blocks within a stage, defined as a sequence of residual blocks with constant dimensionality and no downsampling or upsampling, operate on the same underlying features and refine them iteratively. Substantial changes in representation occur primarily at stage transitions, while representations remain close to the stage input within a stage. Distinguishing these perspectives requires some ability to interpret the latent space representation.

A simple approach to gaining insight into internal model progression is the logit lens method, originally applied to large language models \citep{nostalgebraist_interpreting_2020}. In this method, intermediate latent states are passed directly to the model’s decoder, and the resulting logit distributions are interpreted as the model’s current belief about the next predicted token. By applying this procedure across model depth, it becomes possible to track how predictions evolve as additional processing layers are incorporated. 
In a sense, this approach is analogous to quantum mechanics, where latent states are not interpreted directly but are instead mapped to observable quantities through an operator, in this case the model’s decoder.

An important limitation of applying the logit lens to intermediate states is a potential lack of representational stability across model depth. Intermediate latent states do not necessarily share the same basis as the final state expected by the decoder. Deep models do not possess a privileged basis and individual layers may apply linear transformations, such as rotations, in feature space that are undone later \citep{elhage_mathematical_2021,belrose_eliciting_2023}. While such transformations preserve information, they can render intermediate decodings uninterpretable. 
A potential solution for this is the tuned lens, introduced by \citet{belrose_eliciting_2023}. This method makes use of translators, which are trained affine transformations that align intermediate latent states with the decoder’s input basis. 

In this paper, we adapt the method of \citet{belrose_eliciting_2023} to decode the intermediate states of AI weather models to physical variables. Translators are trained to map from the latent space representation after each processor layer to the representation that is input to the decoder. The output of the decoder is an atmospheric state in the same form as the final forecast. By creating predictions that use only the early processor blocks, we probe the model's current belief, which can be seen as the best possible prediction the model can make at each latent stage. Since this progressively evolving prediction provides a proxy for internal model progression, we aim to gain insight into how forecasts are formed within the latent space and, in doing so, contribute to the development of a general framework for how AI weather models generate predictions. 

Since the essence of this method is to decode intermediate states to physical variables, we use the name \textit{intermediate decoding}. Translators are trained for three AI weather models: \textit{GraphCast}, \textit{Aurora} and \textit{ArchesWeather}, to see how the forecast progression depends on model architecture and size. We then examine how physical fields such as geopotential and wind evolve with processor depth. This allows us to identify which stages of the processing are responsible for different aspects of the forecast including spatial scales and dynamic and thermodynamic properties. 

In the following sections we describe the method of intermediate decoding in detail, and show that it is able to give consistent representations of the model state across processor blocks. Most importantly, we identify a number of fundamental behaviors that appear to be universal among all the models that were examined:
\begin{enumerate}
\item Each processor block successively improves the forecast, with different physical quantities converging toward their final values at different rates.
\item The encoder and decoder do not merely transform physical variables to and from the latent space, they also play a significant role in producing the forecast.
\item The models tend to modify large spatial scales early in processing, moving toward smaller scales with increasing processor depth.
\item Geostrophic balance \citep{bonavita_limitations_2024} holds for synoptic scales of motion at all stages of processing, while the mesoscales come into balance in later layers. 
\end{enumerate}
These behaviors are interesting in themselves, but also because they hint that the different models may be learning a universal underlying representation of the dynamics of the atmosphere.

\section{Methods}\label{methods}

In this study, we apply the method of intermediate decoding to three different models: GraphCast, Aurora, and ArchesWeather. GraphCast is based on a graph neural network architecture, while Aurora and ArchesWeather employ transformer-based architectures \citep{lam_graphcast_2023, bodnar_aurora_2024, couairon_archesweather_2024}. Comparing models with fundamentally different architectures allows us to determine which behaviors are universal and which are model-specific. We first provide an overview of the structures of the three models, followed by a description of the implementation and training of the translators.

\subsection{Models}

\begin{figure}
    \centering
    \def\svgwidth{\linewidth}
    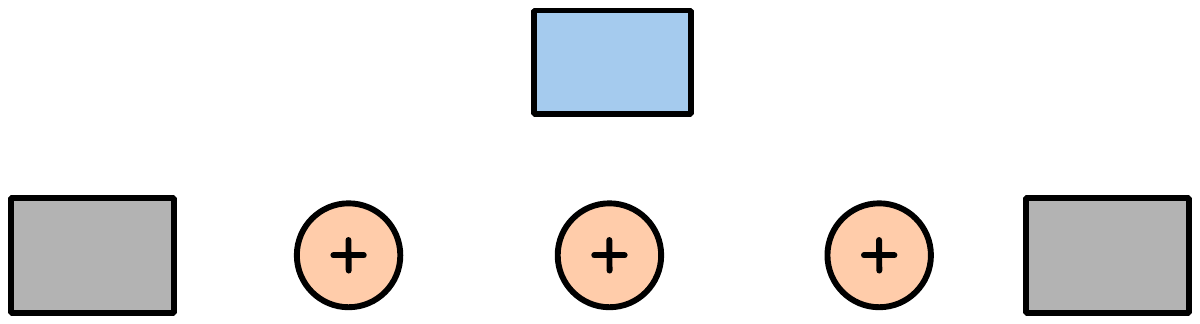
    \vspace{-9.5cm}
    \caption[Residual network architecture]{Residual network architecture. The encoder maps the input to the latent state~$\tilde{X}_0$, which is updated by $L$~processor blocks, each adding an increment~$F_i(\tilde{X}_i)$, creating the intermediate states~$\tilde{X}_i$. The decoder maps the final latent state~$\tilde{X}_L$ to the forecast.}
    \label{Figure:resnet}
\end{figure}

All three models employ a residual network architecture consisting of an encoder, a sequence of processor blocks, and a decoder (Figure~\ref{Figure:resnet}). For encoding, GraphCast uses a graph-based encoder that projects the inputs onto a hierarchical icosahedral mesh with a single vertical level, whereas Aurora and ArchesWeather retain a vertical dimension in the latent state, with four and eight levels, respectively. Aurora uses an attention-based 3D Perceiver encoder, and ArchesWeather uses a convolutional encoder. The resulting latent state $\tilde{X}_0$ is passed through a sequence of processor blocks before decoding. Each block implements a learned function that adds an increment to the current latent state:
\begin{align}
    \tilde{X}_{i+1} = \tilde{X}_i + F_i(\tilde{X}_i),
\end{align}
where $\tilde{X}_i$ denotes the latent state at model depth $i$, and $F_i(\tilde{X}_i)$ represents the increment added by the $i$-th processor block. 

GraphCast consists of $L = 16$, Aurora of $L = 48$, and ArchesWeather of $L = 32$ processor blocks, yielding $L + 1$ latent states per model, given by $\{\tilde{X}^{\mathrm{GC}}_0, \dots, \tilde{X}^{\mathrm{GC}}_{16}\}$, $\{\tilde{X}^{\mathrm{Au}}_0, \dots, \tilde{X}^{\mathrm{Au}}_{48}\}$, and
$\{\tilde{X}^{\mathrm{Ar}}_0, \dots, \tilde{X}^{\mathrm{Ar}}_{32}\}$, where the superscripts $\mathrm{GC},\mathrm{Au},\mathrm{Ar}$ denote latent states of GraphCast, Aurora, and ArchesWeather, respectively. Note that in GraphCast, the latent state is carried jointly by edge and mesh nodes, as edges are also designed as a residual stream; however, we consider only the mesh nodes, as discussed in Section~\ref{methods}\ref{intermediate_decoding_methods}. In GraphCast, processor blocks correspond to message-passing steps on a graph, incorporating both short-range and sparser long-range communication at each step, whereas in Aurora and ArchesWeather, processor blocks are shifted-window (Swin) transformers consisting of an attention layer followed by an MLP. We do not distinguish between these formulations and refer to them uniformly as processor blocks.

\begin{figure*}
    \centering
    \def\svgwidth{\textwidth}
    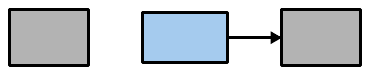

    \vspace{-15cm}
    \caption{Architectural overview of GraphCast, Aurora, and ArchesWeather. Gray boxes denote the encoder and decoder, blue boxes groups of processor blocks, with the corresponding latent states~$\tilde{X}_i$ indicated above each group and the dimensions of the latent state below (for Aurora and ArchesWeather: vertical levels $\times$ latitude $\times$ longitude $\times$ channels; for GraphCast: mesh nodes $\times$ channels). GraphCast operates at constant resolution, whereas Aurora and ArchesWeather downscale and upscale between stages, changing both spatial resolution and channel dimension. Skip connections pass the output of the first stage to later stages, either by addition ($+$) or by concatenation along the channel dimension (conc). For Aurora, the skip connections are shown as implemented in model code version~v1.8.0 (see text).}
    \vspace{-0.1cm}
    \label{Figure:architecture}
\end{figure*}

While GraphCast maintains a constant spatial grid and latent dimensionality throughout processing, Aurora and ArchesWeather employ multi-stage U-Net-style architectures, in which down- and upsampling operations between stages change the spatial resolution and channel dimensionality. These operations are implemented as linear projections that map the channels of the previous stage to a channel dimension of different size.
An overview of the model architectures is provided in Figure~\ref{Figure:architecture}. For Aurora, we show the skip connections as implemented in version v1.8.0 of the model code, which differ from the backbone figure in \cite{bodnar_aurora_2024}: the output of encoding stage~1 is used twice, once as an additive skip connection to the input of decoding stage~3 and once concatenated to the output of decoding stage~3.

All three models take two input timesteps expressed in physical variables, $X^t$ and $X^{t-1}$, along with auxiliary information, including a land-sea mask, orography, and some form of time and positional information. Aurora and ArchesWeather additionally take soil type, while GraphCast includes explicit radiative forcing. GraphCast and ArchesWeather are trained to predict the residual $Y^{t+1}$ between the target state $X^{t+1}$ and the most recent input $X^t$, with the final prediction obtained by adding this residual to the input state,
\begin{align}
X^{t+1} = \mathrm{Model}(X^t, X^{t-1}) = X^t + Y^{t+1}.
\end{align}
Aurora, in contrast, directly predicts the full state $X^{t+1}$. In GraphCast and Aurora, $\Delta t = 6\,\mathrm{h}$, while in ArchesWeather, $\Delta t = 24\,\mathrm{h}$.

GraphCast is evaluated in three configurations: the version presented in the paper at $0.25^\circ$ resolution, referred to as \textit{std}; an operational version at $0.25^\circ$ resolution, additionally fine-tuned on IFS HRES data, referred to as \textit{op}; and a smaller variant at $1^\circ$ resolution using an internal icosahedral mesh with five refinement levels, corresponding to 10{,}242 instead of 40{,}962 nodes, referred to as \textit{small}. All variants are pretrained on ERA5. For all results shown in Section~\ref{sec:results}, the GraphCast versions are qualitatively consistent, and we show the results for the \textit{std} version. For results that are averaged over several intermediate decodings, the \textit{op} and \textit{small} versions were evaluated for a single date only. Aurora operates on a $0.25^\circ$ input and output grid; we use the fine-tuned version, pretrained on ERA5 and subsequently fine-tuned on IFS HRES data. ArchesWeather is trained on ERA5 with $1.5^\circ$ input and output resolution; we use the ArchesWeather-M variant consisting of 32 processor blocks. 

The models differ considerably in scale and complexity. GraphCast comprises 36.7M parameters, Aurora 1.3B, and ArchesWeather-M 89M. Parameter count alone, however, is not a reliable indicator of model complexity: GraphCast and Aurora operate at substantially higher internal resolution and channel dimensionality than ArchesWeather, and also produce higher-resolution outputs, which is reflected in their respective training costs. GraphCast was trained for 896 TPU-days, Aurora for 600 A100-days, and ArchesWeather for only 5 A100-days. Despite the differences in parameter count and hardware, GraphCast and Aurora required a comparable order of magnitude of total compute, with GraphCast consuming roughly 30\% more, whereas ArchesWeather required roughly two orders of magnitude less compute than either.

All three models predict upper-air fields (geopotential $z$, temperature $t$, specific humidity $q$, and wind components $u$, $v$, $w$) as well as surface variables (2\,m temperature $t2m$, 10\,m winds $10u$, $10v$, and mean sea-level pressure $msl$). GraphCast additionally predicts 6\,h accumulated precipitation $tp$ as an extra surface variable. For more detailed descriptions of the models, we refer to the original publications \citep{lam_graphcast_2023, bodnar_aurora_2024, couairon_archesweather_2024}.

\subsection{Intermediate decoding} \label{intermediate_decoding_methods}

The goal of intermediate decoding is to extract the latent space vector from the residual stream after each processor block and send it to the model's decoder to convert to physical variables. It is not sufficient to simply ablate the following processor blocks and send the latent vector directly to the decoder, for two reasons. First, because the residual stream accumulates the updates of all preceding blocks, it is not centered at zero but carries a growing bias. Second, there is no privileged basis in the latent space, so intermediate states need not share the basis expected by the decoder. Here we adopt the solution of \citet{belrose_eliciting_2023} and learn a set of mappings, called \textit{translators}, that take the intermediate representation and translate it to the one that the decoder expects. A translator is defined as a single affine transformation acting on $\tilde{X}_i$, the latent state after processor layer $i$,
\begin{equation}
\tilde{X}_F = W_i \tilde{X}_i + b_i,
\end{equation}
where $W_i$ is a learned linear transformation, interpretable as a change-of-basis matrix that addresses the second issue, and $b_i$ is a learned bias term, corresponding to a change of origin that addresses the first. This maps from an intermediate latent state $\tilde{X}_i$ to the final latent state space $\tilde{X}_F$. Translators are location-independent; the same matrix and bias term are applied at all spatial locations.

The concept of a translator is simple, but the implementation is complicated by details of the model architectures. A first complication is that the upscaling and downscaling layers in Aurora and ArchesWeather change both the number of mesh points and the latent space dimension. A second complication is the presence of other pathways for information to propagate in parallel to the residual stream on mesh points. GraphCast propagates information on edges as well as mesh nodes; however this pathway can be ignored, since this information is not used by the decoder, and would only influence the mesh nodes through the processor layers that are being bypassed. In contrast, the skip connections across multiple blocks in Aurora and ArchesWeather (see Fig.~\ref{Figure:architecture}) cannot be ignored, since they directly modify the state seen by the decoder. The remainder of this subsection explains how the translators are implemented for each model and how they are trained.

GraphCast maintains a latent representation with fixed spatial resolution and channel dimensionality throughout processing, so that intermediate latent states can be decoded directly or mapped by a translator into the final latent space prior to decoding:
\begin{align}
\hat{X}^{t+1}_{i,\,\text{direct}} &= \mathrm{dec}\!\left( \tilde{X}^{\text{GC}}_i \right),
\label{eq:gc_direct} \\[0.8em]
\hat{X}^{t+1}_{i,\,\text{tuned}} &= \mathrm{dec}\!\left(
W^{\rightarrow\tilde{X}^{\text{GC}}_{16}}_{i}\,\tilde{X}^{\text{GC}}_i
+ \mathbf{b}^{\rightarrow\tilde{X}^{\text{GC}}_{16}}_{i}
\right),
\label{eq:gc_tuned} \\[0.8em]
\text{for } & i = 0,\dots,16. \nonumber
\end{align}
Here, $\hat{X}^{t+1}_{i,\,\text{direct}}$ and $\hat{X}^{t+1}_{i,\,\text{tuned}}$ denote the forecasts obtained by decoding $\tilde{X}^{\text{GC}}_i$ directly or after applying the translator $\big(W^{\rightarrow\tilde{X}^{\text{GC}}_{16}}_{i}, \mathbf{b}^{\rightarrow\tilde{X}^{\text{GC}}_{16}}_{i}\big)$, where the arrow superscript denotes the latent state the translator was trained to map to, here $\tilde{X}^{\text{GC}}_{16}$. For $i = 16$, the translator reduces to the identity, i.e., $W^{\rightarrow\tilde{X}^{\text{GC}}_{16}}_{16} = I$ and $\mathbf{b}^{\rightarrow\tilde{X}^{\text{GC}}_{16}}_{16} = \mathbf{0}$. This yields 16 learned translators in total. We use analogous notation for Aurora and ArchesWeather.

The latent representations in Aurora and ArchesWeather vary in both spatial resolution and channel dimensionality within the processing pipeline, making the design of translators more involved. For latent states that share the same spatial resolution as the final latent state, tuned decoding follows the procedure in Equation~(\ref{eq:gc_tuned}), with the modification that the translators learn a channel-wise mapping from 512 to 1024 channels in Aurora and from 192 to 384 channels in ArchesWeather. For direct decoding, we proceed as in equation~(\ref{eq:gc_direct}) and pad missing channels with zeros to match the dimensionality of the final latent space. For downsampled latent states, we apply a sequence of consecutive translators, whose number depends on the number of native upsampling blocks separating the state from the final latent space. In ArchesWeather, which has only one native down- and upsampling block, decoding the latent states $\tilde{X}^{\text{Ar}}_5$--$\tilde{X}^{\text{Ar}}_{28}$ therefore involves two translators. The first maps the intermediate latent state to the state immediately preceding the model's native upsampling block. The native upsampling block is then applied, and a second translator projects the result into the final latent state. This gives:
\begin{align}
\hat{X}^{t+1}_{i,\,\text{direct}} &= \mathrm{dec}\!\left(
\mathrm{con}\big(\mathrm{up}(\tilde{X}^{\text{Ar}}_i),\,\tilde{X}^{\text{Ar}}_{4}\big)
\right),
\label{eq:Ar_direct} \\[0.5em]
\hat{X}^{t+1}_{i,\,\text{tuned}} &= \mathrm{dec}\!\left(
W^{\rightarrow\tilde{X}^{\text{Ar}}_{32}}_{i}\,
\mathrm{con}\big(\mathrm{up}(z^{\text{Ar}}_{i}),\,\tilde{X}^{\text{Ar}}_{4}\big)
+ \mathbf{b}^{\rightarrow\tilde{X}^{\text{Ar}}_{32}}_{i}
\right),
\label{eq:Ar_tuned} \\[0.5em]
z^{\text{Ar}}_{i} &= W^{\rightarrow\tilde{X}^{\text{Ar}}_{28}}_{i}\,\tilde{X}^{\text{Ar}}_i
+ \mathbf{b}^{\rightarrow\tilde{X}^{\text{Ar}}_{28}}_{i},
\qquad i = 5,\dots,28. \nonumber
\end{align}
Here, $\mathrm{up}()$ denotes the model's native upsampling block and $\mathrm{con}()$ denotes channel-wise concatenation with $\tilde{X}^{\text{Ar}}_{4}$, the output of the first encoder layer, which is carried forward via a skip connection. For $i = 28$, the first translator reduces to the identity, i.e., $W^{\rightarrow\tilde{X}^{\text{Ar}}_{28}}_{28} = I$ and $\mathbf{b}^{\rightarrow\tilde{X}^{\text{Ar}}_{28}}_{28} = \mathbf{0}$, so that $z^{\text{Ar}}_{28} = \tilde{X}^{\text{Ar}}_{28}$. 

In Aurora, states in encoder layer~2 and decoder layer~2 are separated from the final latent space by a single native upsampling block, so we proceed as in ArchesWeather, with two translators. States $\tilde{X}^{\text{Au}}_{17}$--$\tilde{X}^{\text{Au}}_{32}$, however, are separated from the final latent space by two native upsampling blocks and therefore require three consecutive translators. For brevity, we give the formulas below only for the one-upsampling-block case:
\begin{align}
\hat{X}^{t+1}_{i,\,\text{direct}} &= \mathrm{dec}\!\left(
\mathrm{con}\big(\mathrm{up}(\tilde{X}^{\text{Au}}_i) + \tilde{X}^{\text{Au}}_{6},\,\tilde{X}^{\text{Au}}_{6}\big)
\right),
\label{eq:Au_direct} \\
\hat{X}^{t+1}_{i,\,\text{tuned}} &= \mathrm{dec}\!\Big(
\mathrm{con}\big(W^{\rightarrow\tilde{X}^{\text{Au}}_{48}}_{i}\,\mathrm{up}(z^{\text{Au}}_{i})
+ \mathbf{b}^{\rightarrow\tilde{X}^{\text{Au}}_{48}}_{i}  \nonumber \\
&\qquad\ + \tilde{X}^{\text{Au}}_{6}, \tilde{X}^{\text{Au}}_{6}\big)
\Big),
\label{eq:Au_tuned} \\
z^{\text{Au}}_{i} &= W^{\rightarrow\tilde{X}^{\text{Au}}_{42}}_{i}\,\tilde{X}^{\text{Au}}_i
+ \mathbf{b}^{\rightarrow\tilde{X}^{\text{Au}}_{42}}_{i}, \nonumber
\\[0.8em]
\text{for } \;\; i &= 7,\dots,16\;\text{and}\;i = 33,\dots,42. \nonumber
\end{align}
Here, $\tilde{X}^{\text{Au}}_{6}$ is the output of the first encoder layer, which enters both through the additive skip connection after upsampling and through the final concatenation. For $i = 42$, the first translator reduces to the identity, i.e., $W^{\rightarrow\tilde{X}^{\text{Au}}_{42}}_{42} = I$ and $\mathbf{b}^{\rightarrow\tilde{X}^{\text{Au}}_{42}}_{42} = \mathbf{0}$, so that $z^{\text{Au}}_{42} = \tilde{X}^{\text{Au}}_{42}$.

Note that our intermediate decoding method retains the up- and downsampling blocks as well as the skip connections, and only translates within stages of constant dimensionality. In principle, it would be simpler to learn a translator that maps directly from the current to the final latent state. However, as shown in the supplementary material, this performs worse: translating directly ignores the high-resolution information carried by the skip connections, resulting in a less consistent decoded state.

Translators are trained on latent states extracted from forward passes of the models for the year 2020. For GraphCast, we use 1462 samples, obtained by running 6\,h forecasts over the full year, split into 1170 training and 292 validation samples. For Aurora, we use 1464 samples analogously obtained, split into 1172 training and 292 validation samples. For ArchesWeather, the corresponding data is obtained by running 24\,h forecasts, yielding 1165 training and 291 validation samples. All models are initialized such that the forecasts are valid at 00:00, 06:00, 12:00, and 18:00 UTC. Training minimizes the mean squared error (MSE) between between each translated latent state $W_i \tilde{X}_i + b_i$ and its target latent state using the Adam optimizer with a learning rate of $5 \times 10^{-4}$, and weight matrices are initialized as the identity. For GraphCast, Aurora and ArchesWeather, we use a batch size of 2, where each batch consists of full global fields, resulting in 585, 586, and 583 weight updates per epoch, respectively. An overview of convergence behavior is provided in Table~\ref{tab:convergence}. Full training curves are available in the supplementary material.

\begin{table}[h]
\centering
\caption{Epochs required for translators to converge to within 1\% and 0.1\% relative improvement in validation loss (range and median).}
\label{tab:convergence}
\resizebox{\columnwidth}{!}{%
\begin{tabular}{lcc}
\toprule
\textbf{Model} & \textbf{1\% convergence (epochs)} & \textbf{0.1\% convergence (epochs)} \\
\midrule
GraphCast              & 4--7 (median 5)   & 6--13 (median 10) \\
Aurora                 & 2--18 (median 2)  & 2--26 (median 4)  \\
ArchesWeather & 2--7 (median 3.5) & 2--14 (median 7)  \\
\bottomrule
\end{tabular}}
\end{table}

\section{Results} \label{sec:results}
\subsection{Representational stability}\label{subsection:represstab}
Before examining the decoded latent states, we evaluate how well the translators correct for \textit{representational drift}, i.e., changes in how features are expressed across the channels of the latent state with model depth.

In this paper, a \emph{channel} refers to a single dimension of the latent vector, whereas a \emph{feature} refers to a property of the state that is meaningful to the model and may be encoded along an arbitrary direction in latent space, not necessarily aligned with a single channel. When we speak of a \emph{representation}, we mean the collection of features encoded in a latent state, expressed in an arbitrary latent basis. In \cite{olah_zoom_2020}, a representation is described as the vector space formed by the activations of all neurons in a layer. 

Under a change of basis, the same underlying features are represented by different vectors. Consequently, changes in latent state vectors across model depth may indicate that the same features are represented in a different basis, rather than that the model genuinely alters the features it represents. In the first case, the translators will perform the necessary change of basis, producing a high degree of correlation between the representations at different layers, even if this was not present in the untranslated representation.

\begin{figure*}[!t]
    \centering
    \includegraphics[width=1\linewidth]{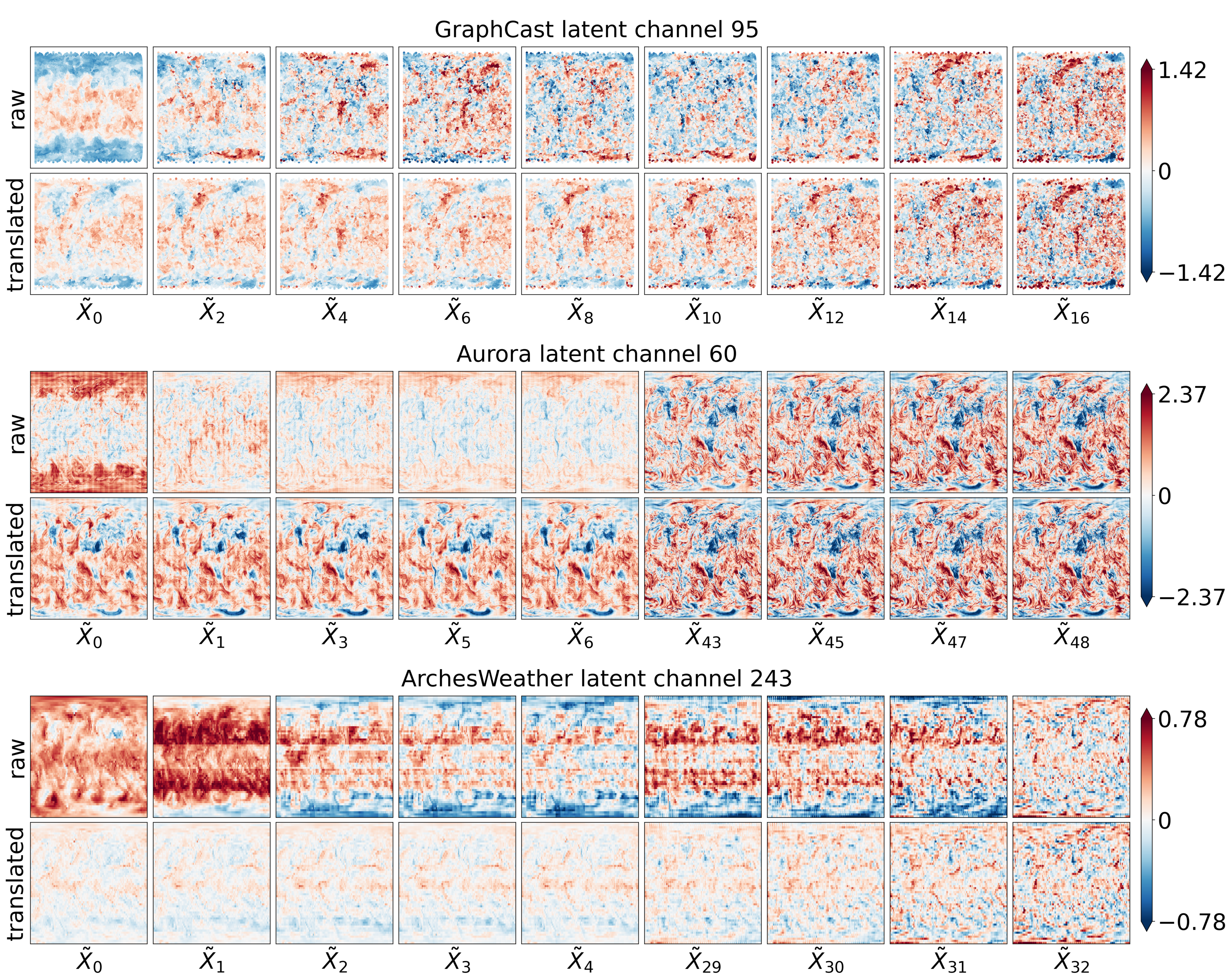}
    \caption{Progression of extracted latent space channels across model depth for a forecast valid at 12:00 UTC on 10 March 2020. Shown is the progression of a representative channel for each model on the respective internal model grid, without and with the use of translators. For GraphCast, every second latent state on the icosahedral mesh is shown. For Aurora and ArchesWeather, latent states from the first and last layer are shown on the lat/lon grid on a single vertical level. Colorbar limits are set to the 99th percentile of the absolute raw latent values over all shown blocks of each model.}
    \label{fig:latent_channels}
\end{figure*}

Figure~\ref{fig:latent_channels} illustrates representational drift for one representative channel per model, with and without applying the trained translators. Without translators, the spatial pattern of a channel changes abruptly across depth, indicating that the channel is associated with different features at different depths. Among the many channels we inspected (not shown), the degree of drift varies, with some channels changing completely, while others developing more slowly. We show channels from the middle of this range, which exhibit both abrupt changes and gradual development. For Aurora and ArchesWeather, we compare the first and last layer, which share the same spatial resolution and are connected by skip connections; changes across the down- and upscaling blocks are expected, since these blocks mix channels.

With translators, the abrupt changes disappear and the channels progress more smoothly, successively adding detail to the spatial pattern. A channel either resembles the final latent state from the first latent state onward, or initially appears as a weak pattern that is nonetheless consistent with the final state. This works even though the same translator is applied at all spatial positions. The development across the full depth of Aurora and ArchesWeather is shown in the supplementary material for a selection of channels.

\begin{figure*}[!t]
    \centering
    \includegraphics[width=1\linewidth, trim=0 0 0 0, clip]{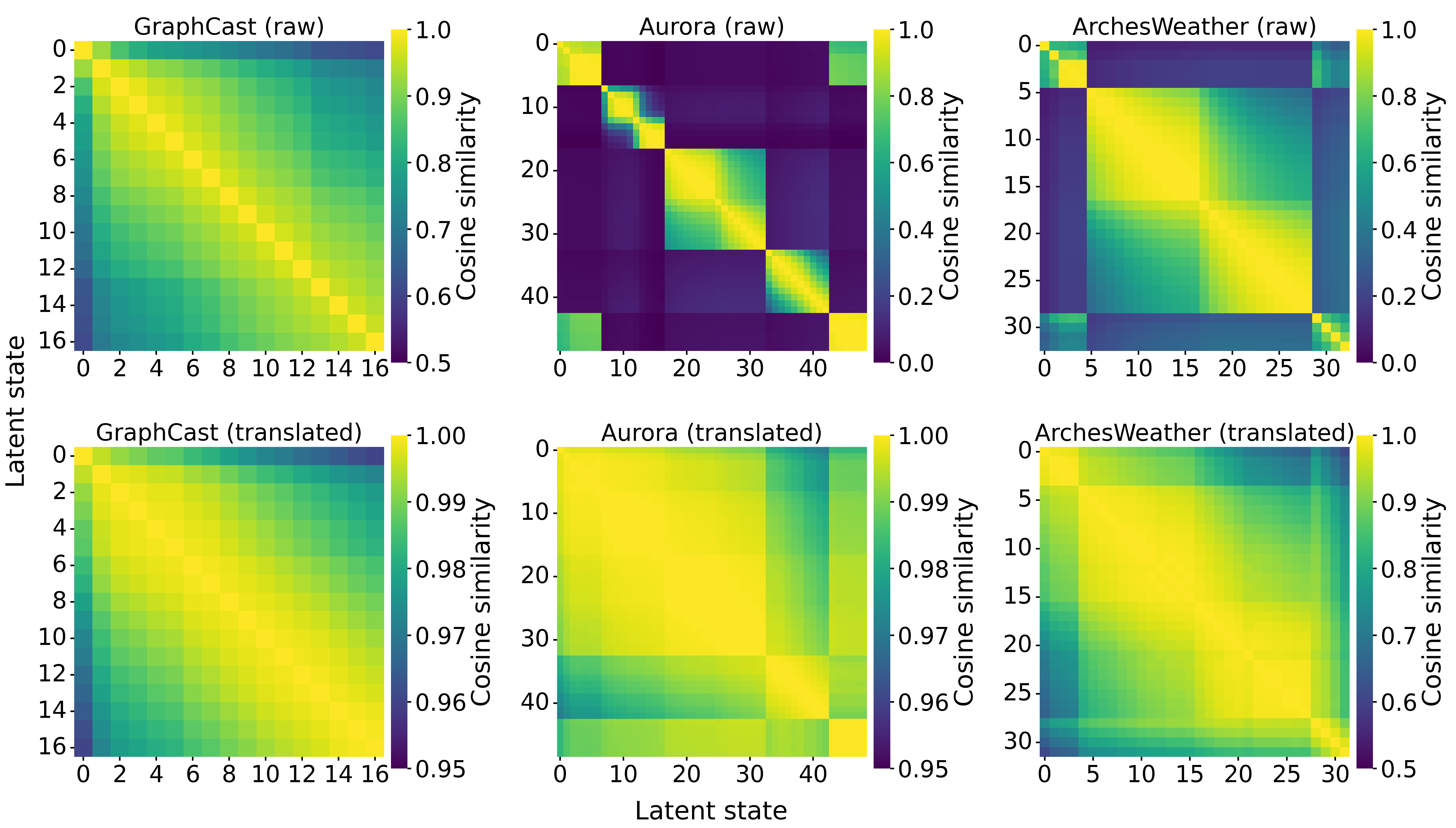}
    \caption{Comparison of covariance structures across model depth using Frobenius cosine similarity, for untranslated (raw) and translated latent states. Covariance matrices are computed from one year of data (2020) for GraphCast, Aurora, and ArchesWeather, and compared across all pairs of latent states; high similarity indicates shared covariance structure. Top panels show covariance similarity for untranslated latent states; bottom panels show covariance similarity for latent states after applying the respective translators.}
    \label{fig:frobenius_cos_sim}
\end{figure*}

To quantify representational drift, we follow \citet{belrose_eliciting_2023} and compare covariance matrices of latent states across model depth. Using the latent-space data from 2020 extracted for translator training, we compute the covariance matrix at each mesh point, average it globally, and compare the resulting matrices across depth using the Frobenius cosine similarity, both with and without translators.

Formally, let $\tilde{\mathbf{x}}_{i,t}^{(n)} \in \mathbb{R}^d$ denote the vector of latent state $\tilde{X}_i$ for sample $t$ at location $n$, with $t=1,\dots,T$ and $n=1,\dots,N$. The sample mean at location $n$ is
\begin{equation}
\boldsymbol{\mu}_i^{(n)} = \frac{1}{T}\sum_{t=1}^T \tilde{\mathbf{x}}_{i,t}^{(n)},
\label{eq:sample_mean}
\end{equation}
and the corresponding sample covariance is given by
\begin{equation}
C_i^{(n)} = \frac{1}{T-1} \sum_{t=1}^T \big(\tilde{\mathbf{x}}_{i,t}^{(n)} - \boldsymbol{\mu}_i^{(n)}\big)\big(\tilde{\mathbf{x}}_{i,t}^{(n)} - \boldsymbol{\mu}_i^{(n)}\big)^\top .
\label{eq:sample_cov}
\end{equation}
Here, $T$ denotes the total number of samples available from one year of inference runs, corresponding to $T=1462$ for GraphCast, $T=1464$ for Aurora, and $T=1456$ for ArchesWeather. The covariance matrix for a complete latent state $\tilde{X}_i$ is obtained by averaging across all $N$ locations,
\begin{equation}
C_i = \frac{1}{N}\sum_{n=1}^N C_i^{(n)} .
\label{eq:avg_cov}
\end{equation}

To analyze the progression across model depth, we compare the covariance matrices of all pairs of latent states using the Frobenius cosine similarity,
\begin{equation}
S_{ij} = \frac{\langle C_i, C_j\rangle_F}{\|C_i\|_F \,\|C_j\|_F},
\qquad i,j = 0,\dots,L,
\label{eq:frob_cos}
\end{equation}
where $\langle\cdot,\cdot\rangle_F$ denotes the Frobenius inner product (the sum over all elementwise products), $\|\cdot\|_F$ the Frobenius norm, and $L$ the total number of processor blocks in the model. 

Comparing covariance matrices requires a common shape, but in Aurora and ArchesWeather the channel dimensionality changes between stages. For untranslated latent states, we therefore zero-pad the channels to the largest dimensionality reached in each model (2048 for Aurora, 384 for ArchesWeather). For translated latent states, we use separately trained translators without skip connections that map directly to the final latent state, which gives the dimensionality of the final latent state (1024 for Aurora, 384 for ArchesWeather).

Figure~\ref{fig:frobenius_cos_sim} shows the similarity of covariance matrices across model depth. Without translators (top row), similarity decreases gradually between latent states with distance, across the full depth of GraphCast and within stages for Aurora and ArchesWeather. This indicates that neighboring states express features in a similar basis. Across stages of Aurora and ArchesWeather, the similarity is low to negligible, which is expected, as down- and upsampling blocks mix channels. Since additive or concatenative skip connections directly connect the first and last layer in Aurora and ArchesWeather, similarity is again stronger between these two layers.

When translators are applied (bottom row), the covariance matrices become considerably more aligned across model depth. This suggests that most of the drift seen without translators reflects a change of basis rather than a change in the underlying features. The alignment is not perfect, and we attribute the remaining misalignment to non-linear changes in the representation, i.e., the model progressively constructing the forecast within the processor. For Aurora, the downscaled latent states align nearly as well as in GraphCast, whereas for ArchesWeather the alignment remains considerably weaker. This may relate to differences in the type and placement of skip connections in the two models, but might also result indirectly from other architectural differences. In both Aurora and ArchesWeather, the stage structure remains visible after translation, consistent with the view of iterative refinement.

\subsection{Evolution of forecast with processor depth}\label{subsection:performance}

\begin{figure*}[!t]
    \centering
    \includegraphics[width=1\linewidth]{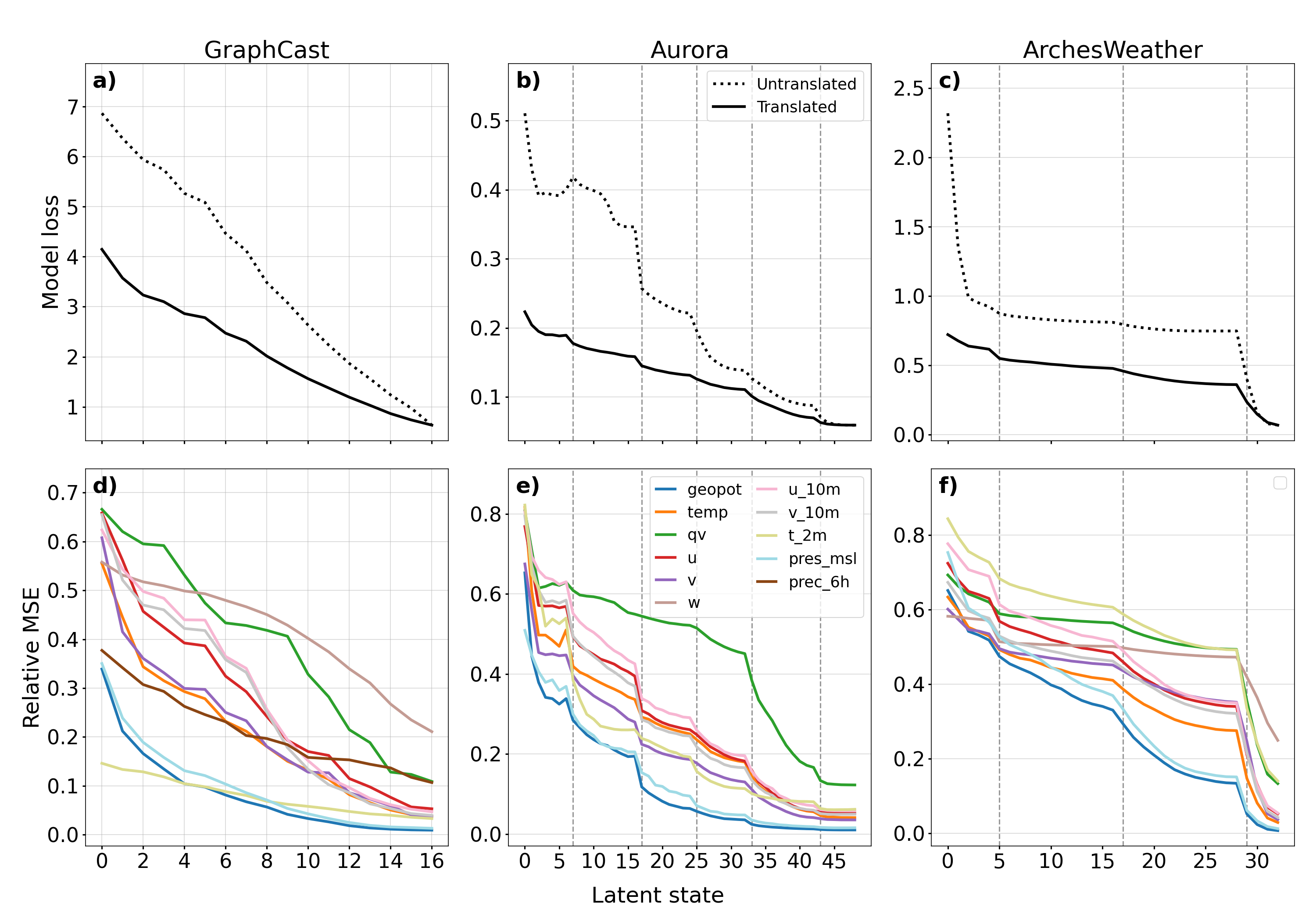}
    \caption{Model loss and relative MSE across latent states for intermediate decoding, averaged over 48 runs (the 10th of each month in 2020 at 00, 06, 12, and 18\,UTC). Panels (a--c) show each model's native loss function for GraphCast, Aurora, and ArchesWeather, respectively, for direct (dotted) and tuned (solid) decodings. Panels (d--f) show the relative MSE of the tuned decodings per variable for the surface variables (\texttt{u\_10m}, \texttt{v\_10m}, \texttt{t\_2m}, \texttt{pres\_msl}, and, for GraphCast only, \texttt{prec\_6h}) and the variables on the 500\,hPa level (\texttt{geopot}, \texttt{temp}, \texttt{qv}, \texttt{u}, \texttt{v}, and \texttt{w}). For Aurora and ArchesWeather, dashed lines mark the first latent state of a new stage.}
    \label{fig:mse_and_loss}
\end{figure*}

We assess the forecast performance of the intermediate decodings by sending intermediate latent states to the decoder. We evaluate each model's native loss function, as well as the relative mean squared error (MSE) per variable, for 48 runs across 2020. The results, averaged over all runs, are shown in Figure~\ref{fig:mse_and_loss}: panels a--c show the loss for direct (dotted) and tuned (solid) decodings, and panels d--f the relative MSE per variable for tuned decodings. 

Tuned decoding yields a substantially lower loss than direct decoding at all depths and for all models (Figure~\ref{fig:mse_and_loss}a--c), consistent with the representational drift shown in Section~\ref{sec:results}\ref{subsection:represstab}. Spatial maps of direct decodings (not shown) additionally exhibit artifacts reflecting the internal model structure (e.g., attention windows or graph structure) and unphysically large values for individual variables, particularly in early layers. We therefore do not consider direct decoding further.

For tuned decodings, the mean loss decreases with each additional processor block in all models, with the exception of the last block of the first encoder stage in Aurora, which causes a slight increase in all runs. 
The overall decrease is approximately linear in GraphCast and Aurora, while in ArchesWeather, the final stage of four blocks, where the high- and low-resolution processing streams are concatenated, yields a particularly strong loss reduction.
The stage structure of the architectures is clearly visible: in all models, the loss reduction is strongest at the beginning of a stage and becomes progressively smaller toward its end. In GraphCast, which has a single stage, this pattern is roughly true across the entire processor, whereas in Aurora and ArchesWeather it repeats within each stage, so that the loss curve flattens toward the end of each stage. This pattern is consistent with iterative refinement, in which a new representation is established at a stage transition and then refined by subsequent blocks. 

The relative MSE for individual variables is shown in Figure~\ref{fig:mse_and_loss}d--f.
For each variable, we compute the area-weighted MSE of the intermediate decoding $\hat{X}_i^{t+1}$, where $i$ denotes the number of processor blocks included in the forecast, against the ERA5 target $X^{t+1}$, and normalize it by the MSE of a persistence forecast, i.e., the most recent model input $X^t$, which is taken from ERA5:
\begin{align}
\mathrm{RE}_i =
\frac{\mathrm{MSE}(\hat{X}_i^{t+1},X^{t+1})}{\mathrm{MSE}(X^{t},X^{t+1})} \,.
\label{eq:mse_fc}
\end{align}
A value of 1 thus corresponds to the error of a persistence forecast.

As was seen for the total loss, the relative MSE of individual variables decreases with model depth in all models. Notably, the encoder--decoder pair alone (latent state $\tilde{X}_0$) reduces the error more than any single processor block, to about 30--70\% of persistence for most variables in GraphCast, 50--85\% in Aurora, and 60--80\% in ArchesWeather. The encoder and decoder thus contribute substantially to the forecast, which we examine in more detail in the next section.

The relative performance of individual variables is remarkably consistent across models. For instance, geopotential at 500\,hPa and mean sea level pressure are already predicted markedly better than specific humidity at $\tilde{X}_0$ and improve quickly from the earliest layers. Specific humidity, in contrast, improves most strongly toward the final layers. An exception is 2\,m temperature, which is the best-predicted variable at $\tilde{X}_0$ in GraphCast but the worst in Aurora and ArchesWeather. The good initial performance of GraphCast may result from the inclusion of explicit radiative forcing information, which is important for predicting the diurnal cycle.

Overall, it is striking how similar the different models appear in the progression of the forecasts with layer depth. It was not obvious \emph{a priori} that the loss would decrease with nearly every processor block, following a similar stage-wise pattern of diminishing reductions, or that individual variables would follow such similar progressions across markedly different architectures.

\subsection{Encoder--decoder forecast}

\begin{figure*}[!t]
    \centering
    \includegraphics[width=0.85\textwidth]{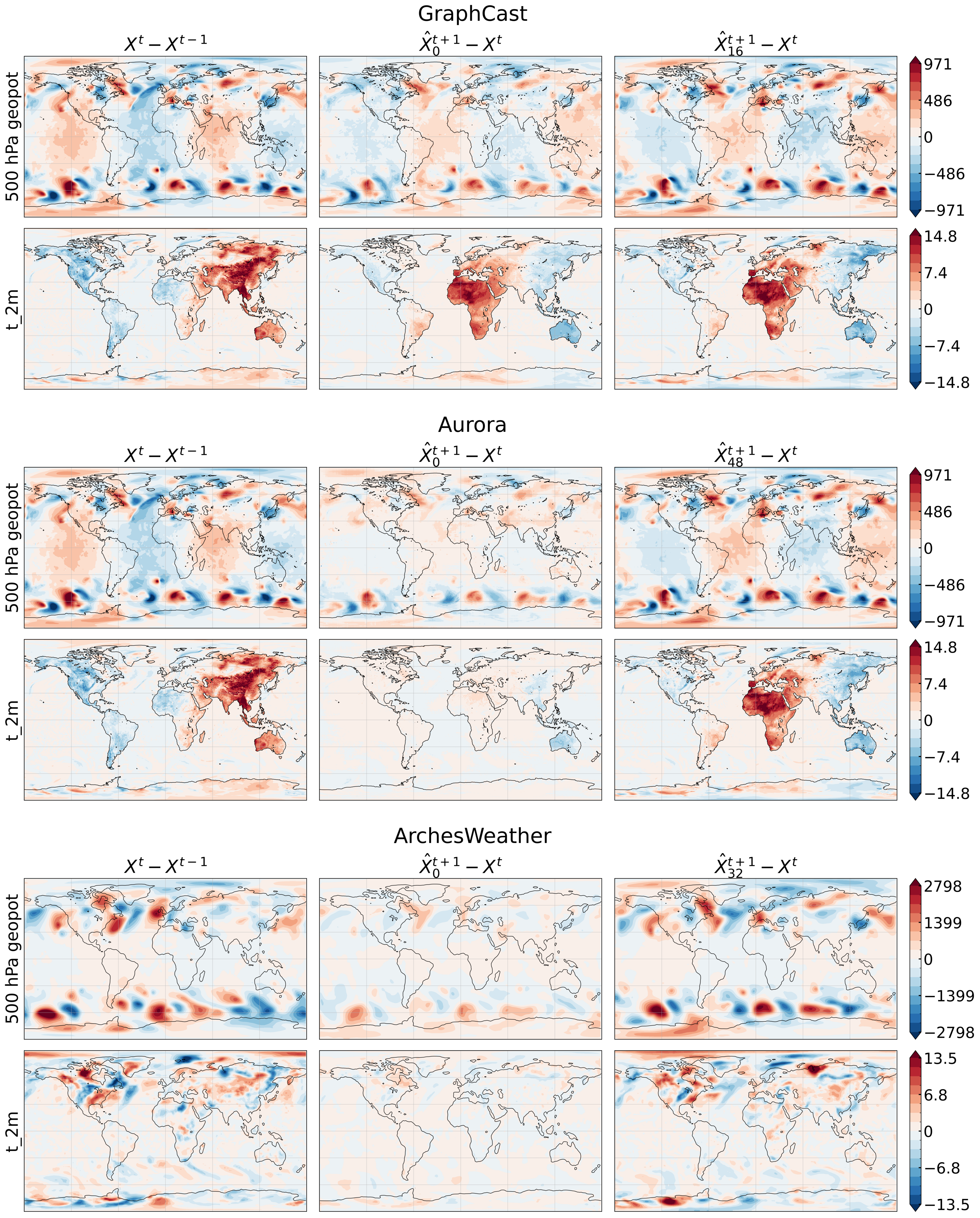}
    \caption{Comparison of encoder--decoder output and full model prediction for GraphCast, Aurora, and ArchesWeather, for a forecast valid at 12:00~UTC on 10 March 2020. All panels show increments relative to the most recent input state: (left) the difference between the two input states, $X^{t}-X^{t-1}$; (middle) the encoder--decoder-only increment, $\hat{X}^{t+1}_0-X^{t}$; and (right) the increment of the full model prediction, $\hat{X}^{t+1}_L-X^{t}$. Shown variables are geopotential at 500\,hPa and 2\,m temperature. Colorbar limits are set to 60\,\%  of the maximum absolute value for each row.}
    \label{fig:encoder_comparison}
\end{figure*}

\begin{figure*}[!t]
    \centering
    \includegraphics[width=0.97\linewidth]{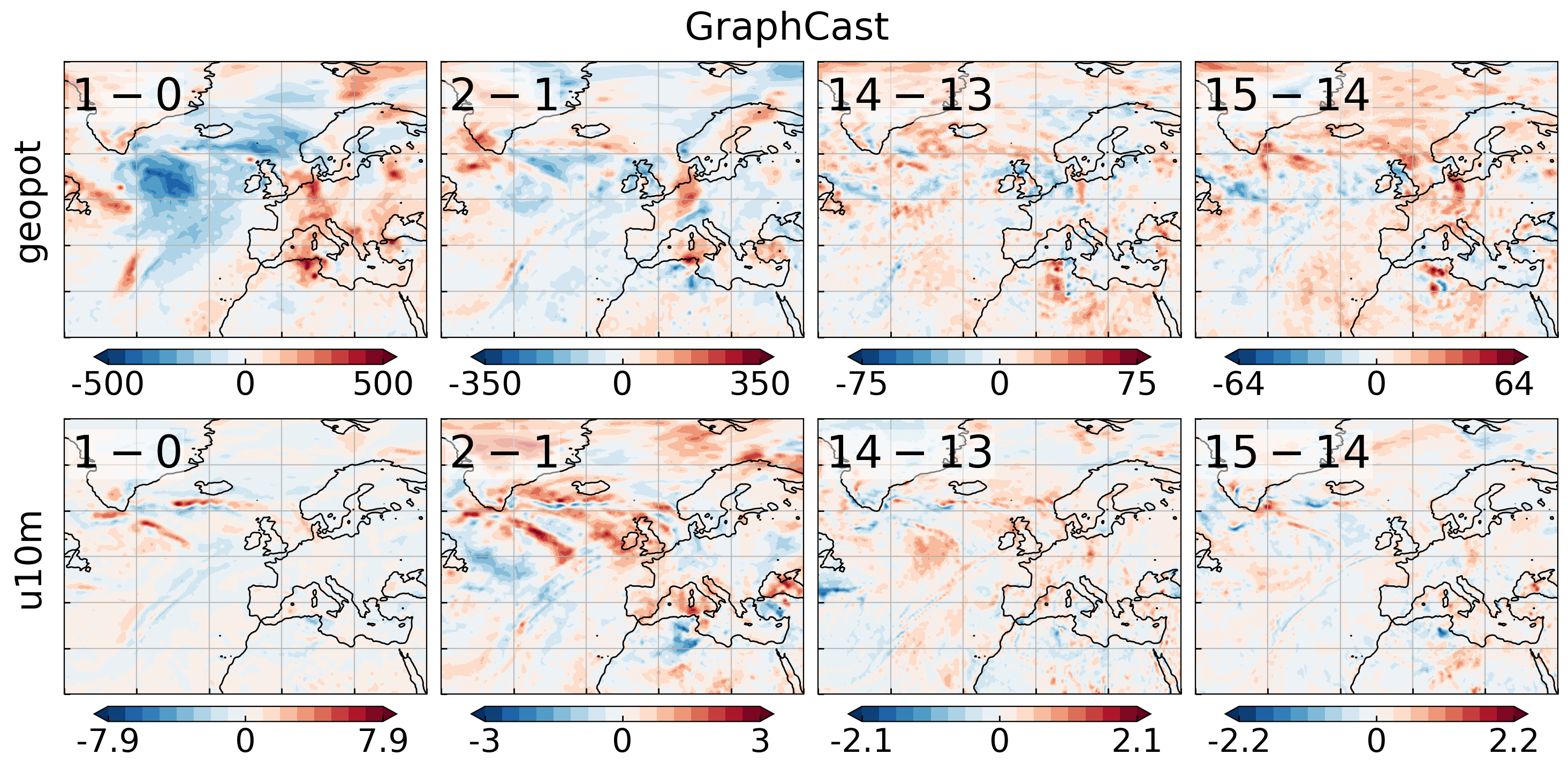}
    \caption{Increments $\Delta\hat{X}^{t+1}_i$ of consecutive intermediate decodings of GraphCast for processor blocks 1, 2, 14, and 15, for the forecast valid at 12:00~UTC on 10 March 2020. Top row: geopotential at 500\,hPa ($\mathrm{m^2\,s^{-2}}$); bottom row: 10\,m zonal wind ($\mathrm{m\,s^{-1}}$). Note the different color scales.}
    \label{fig:gc_increments1}
\end{figure*}

Having established that the encoder--decoder pair produces a substantial loss reduction, we examine the resulting forecast in more detail. This strong reduction is surprising, as without the processor, mesh nodes cannot exchange information and thus cannot aggregate large-scale context. However, for many variables, the encoder--decoder-only forecast already strongly resembles the final prediction, yielding physically meaningful large-scale output. As examples, we present two variables: geopotential at 500\,hPa, which is among the best-predicted variables with the encoder--decoder pair alone, and 2\,m temperature, which differs most strongly across models.

Figure~\ref{fig:encoder_comparison} shows spatial maps of the encoder--decoder output alongside the full model prediction for both variables and all three models, using an example forecast valid at 12:00~UTC on 10 March 2020. To allow a direct comparison, all panels show increments relative to the most recent input state. The left column shows the difference between the two input states as reference, $X^{t}-X^{t-1}$, the middle column the increment produced by the encoder--decoder pair alone, $\hat{X}^{t+1}_0-X^{t}$, and the right column the increment of the full model prediction, $\hat{X}^{t+1}_L-X^{t}$.

For geopotential, the encoder--decoder output captures the large-scale structure of the final prediction in all three models. The fields show clear Rossby wave patterns that are spatially consistent with the full forecast and are shifted downstream compared to the input-difference pattern in the left column. Furthermore, for GraphCast and Aurora, a semidiurnal tidal pattern (12-hour periodicity) is visible for the encoder--decoder output, and is particularly pronounced for GraphCast. To quantify this, we calculate the area-weighted, centered pattern correlation between the encoder--decoder and the full-model increments and take the average over the 48 evaluation runs used in Section~\ref{sec:results}\ref{subsection:performance}. For the 500\,hPa geopotential, the values are 0.82, 0.61, and 0.61 for GraphCast, Aurora, and ArchesWeather, respectively.

For 2\,m temperature, the picture differs across models: in GraphCast, the encoder--decoder output is already close to the full prediction, with a correlation of 0.94, consistent with 2\,m temperature being its best-predicted variable at $\tilde{X}_0$, whereas in Aurora and ArchesWeather it deviates more strongly, with correlations of 0.45 and 0.46, respectively. For the other variables, the correlations range from about 0.45 to 0.86, with the highest values generally found in GraphCast.

It is notable that the models can produce these large-scale patterns before information is exchanged between nodes via processor blocks. This may be enabled by the two input time steps, together with the local spatial context aggregated during encoding and decoding, which yields temporal and spatial gradients of the physical variables. Additionally, all models employ some form of spatial and temporal encoding, so nodes have access to their current position and the time of day and year. This appears sufficient to predict repeating patterns such as Rossby waves, the semidiurnal tide, and the diurnal cycle of surface temperature. Moreover, since the encoder and decoder are trained jointly with the processor rather than in isolation, they directly contribute to the forecast rather than merely preparing the input for later stages. These results show that the encoder--decoder pair does not act as a pair of inverse operations, but instead directly creates physically meaningful, large-scale increments without the processor aggregating spatial context.

\subsection{Intermediate decoding fields and scale progression}\label{subsection:scaleprogression}

\begin{figure*}[!t]
    \centering
    \includegraphics[width=1\linewidth]{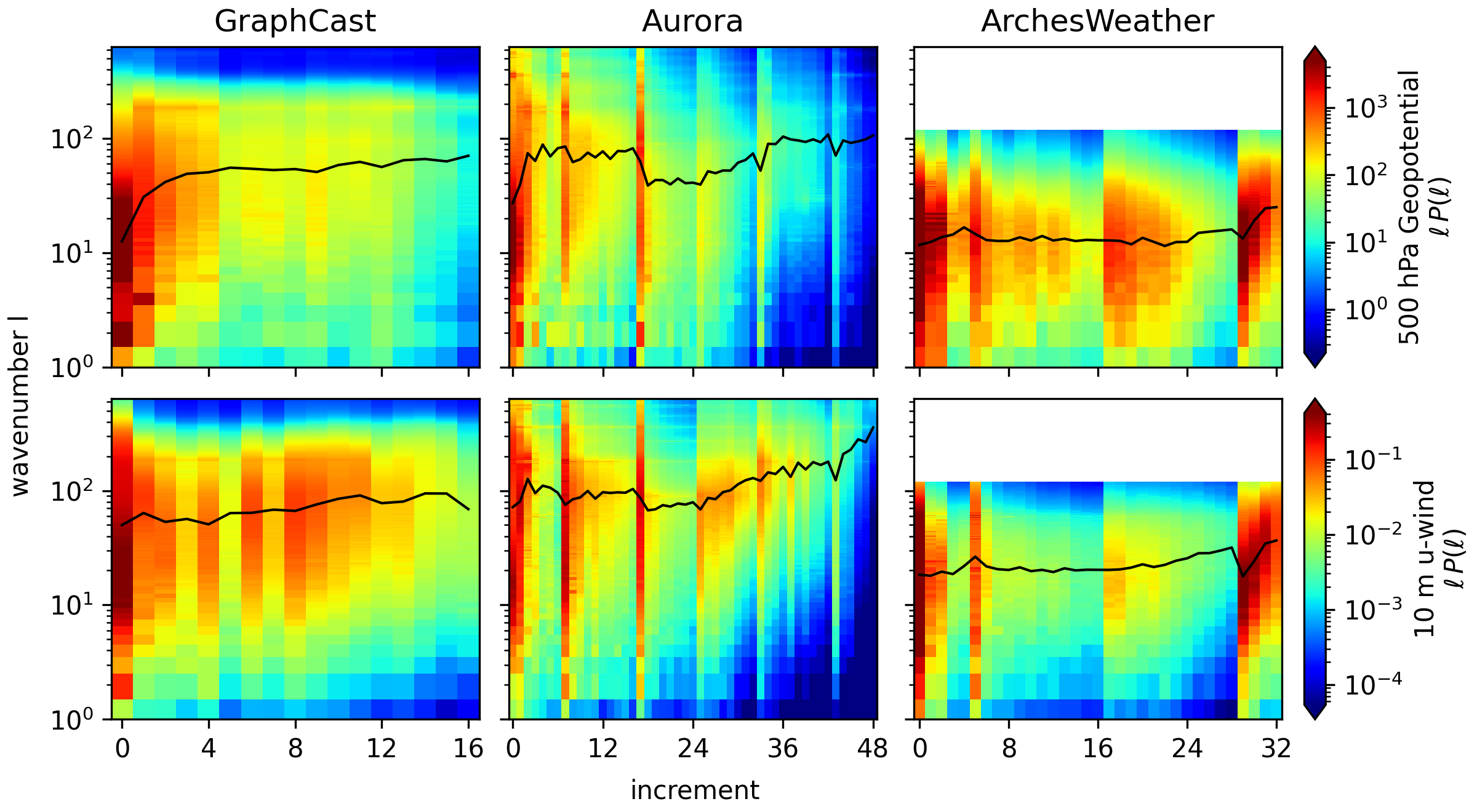}
    \caption{Power spectra $\ell\, P_i(\ell)$ of the increments $\Delta\hat{X}^{t+1}_i$ as a function of total wavenumber $\ell$ and layer depth $i$, for geopotential at 500\,hPa (top) and 10\,m zonal wind (bottom), for GraphCast, Aurora, and ArchesWeather, averaged over the 48 evaluation runs (the 10th of each month in 2020 at 00, 06, 12, and 18\,UTC). Black lines show the spectral centroid of each increment (Eq.~\ref{eq:spectral_centroid}). The first increment ($i=0$) is that of the encoder--decoder-only forecast, $\hat{X}^{t+1}_0 - X^t$. Spectra of ArchesWeather are truncated at $\ell = 120$, corresponding to its $1.5^\circ$ resolution. Colorbar limits are set to the 0.2 and 99.8 percentiles of all values in each row.}
    \label{fig:spectra_layers}
\end{figure*}

The contribution of individual processor blocks is isolated by computing the increment between two consecutive intermediate decodings,
\begin{equation}
    \Delta\hat{X}^{t+1}_i=\hat{X}^{t+1}_i-\hat{X}^{t+1}_{i-1}, \qquad i = 1,\dots,L,
\label{eq:increments}
\end{equation}
with $\Delta\hat{X}^{t+1}_0 = \hat{X}^{t+1}_0 - X^t$ denoting the increment of the encoder--decoder pair. We show results for geopotential at 500\,hPa, representing the large-scale dynamics, and 10\,m zonal wind, a surface variable with pronounced smaller-scale structure such as fronts. Increment maps are shown for the example forecast valid at 12:00~UTC on 10 March 2020.

Figure~\ref{fig:gc_increments1} shows the increments of two early (1 and 2) and two late (14 and 15) processor blocks of GraphCast over Europe and the North Atlantic; the increments of all processor blocks for this region are provided in the supplementary material. For both geopotential at 500\,hPa and 10\,m zonal wind, the early processor blocks produce smooth, large-scale corrections to the forecast, whereas the later blocks add finer-grained adjustments. At the same time, the magnitude of the increments decreases strongly, from up to about $300\,\mathrm{m^2\,s^{-2}}$ in the first blocks to about $40\,\mathrm{m^2\,s^{-2}}$ in the late blocks for geopotential, and from up to about $4.7$ to about $1.3\,\mathrm{m\,s^{-1}}$ for 10\,m zonal wind. This process of iterative refinement can also be seen in the 10\,m zonal wind: elongated increments along the fronts south of Greenland and Iceland already appear in the early processor blocks and are enhanced by several further blocks. Other variables show a similar progression in scale and magnitude, and iterative refinement along fronts is also visible for variables such as 10\,m meridional wind, specific humidity, and precipitation.

For Aurora and ArchesWeather, increment maps for geopotential and 10\,m zonal wind for selected processor blocks within each stage are provided in the supplementary material. Visual inspection is more difficult for these models, as their processors are divided into stages operating at different resolutions. In addition, the increments of both models can exhibit tiling patterns, which appear as a fine checkerboard at the scale of individual tokens in Aurora and as larger patches in ArchesWeather, likely associated with the attention window size. Nonetheless, the increment fields of both models show a progression toward smaller scales and lower magnitudes, both across the full model depth and partly within individual stages. This progression holds across variables. Iterative refinement along fronts is also visible in Aurora, but less clearly in ArchesWeather owing to the coarser model resolution.

To quantify the scale progression, we transform the increments of the decoded fields to spectral space and compute their power spectra $P_i(\ell)$ as a function of total spherical wavenumber $\ell$, together with the corresponding spectral centroids,
\begin{equation}
\mathrm{Centroid}_i
=
\frac{\sum_{\ell} \ell \, P_i(\ell)}
{\sum_{\ell} P_i(\ell)}.
\label{eq:spectral_centroid}
\end{equation}
To avoid aliasing, we truncate the spectra of the $0.25^\circ$ outputs of GraphCast and Aurora at wavenumber 639, which matches the spectral resolution of ERA5, and those of the $1.5^\circ$ outputs of ArchesWeather at wavenumber 120. 

Figure~\ref{fig:spectra_layers} shows the power spectra of the global increment fields for all increments of the three models, again for 500\,hPa geopotential and 10\,m zonal wind. Black lines show the evolution of the corresponding spectral centroids with layer depth. Power spectra and spectral centroids are averaged over the 48 evaluation runs. All models show a progression from large-scale, high-magnitude increments to smaller scales and magnitudes. The spectral centroid increases from the encoder--decoder increment $\Delta\hat{X}^{t+1}_0$ to the final increment $\Delta\hat{X}^{t+1}_{L}$ ($L = 16$, 48 and 32 for GraphCast, Aurora and ArchesWeather, respectively). In GraphCast, the spectral centroid increases from 13 to 71 for geopotential and from 50 to 69 for 10\,m zonal wind, from 27 to 107 and from 72 to 357 in Aurora, and from 12 to 25 and from 18 to 37 in ArchesWeather. This progression is smooth in GraphCast, whereas Aurora and ArchesWeather show a pronounced stage structure. At the beginning of each stage, the magnitude peaks on all scales and subsequently decreases, with a progression to smaller scales within each stage. In Aurora, the scale progression is interrupted by a jump to larger scales at the beginning of encoding stage 3, which acts on the coarsest internal resolution. In ArchesWeather, the last four processor blocks show a strong increase in magnitude, which is consistent with the strong loss reduction in this stage (Section~\ref{sec:results}\ref{subsection:performance}). The spectra for all predicted variables of each model are shown in the supplementary material and exhibit the same patterns.

These results show that the internal forecast construction progresses from large to small scales in all three models. This finding is intuitive, as most of the variance in the atmosphere resides at large scales; to reduce loss substantially, the model would therefore be expected to first adjust large scales before refining smaller ones. This progression is also consistent with findings on internal model increments reported by \citet{craig2026physicsaiweathermodels}.

\subsection{Physical consistency of intermediate decodings}\label{subsec:phybalances}

\begin{figure*}[!t]
    \centering
    \includegraphics[width=1\linewidth]{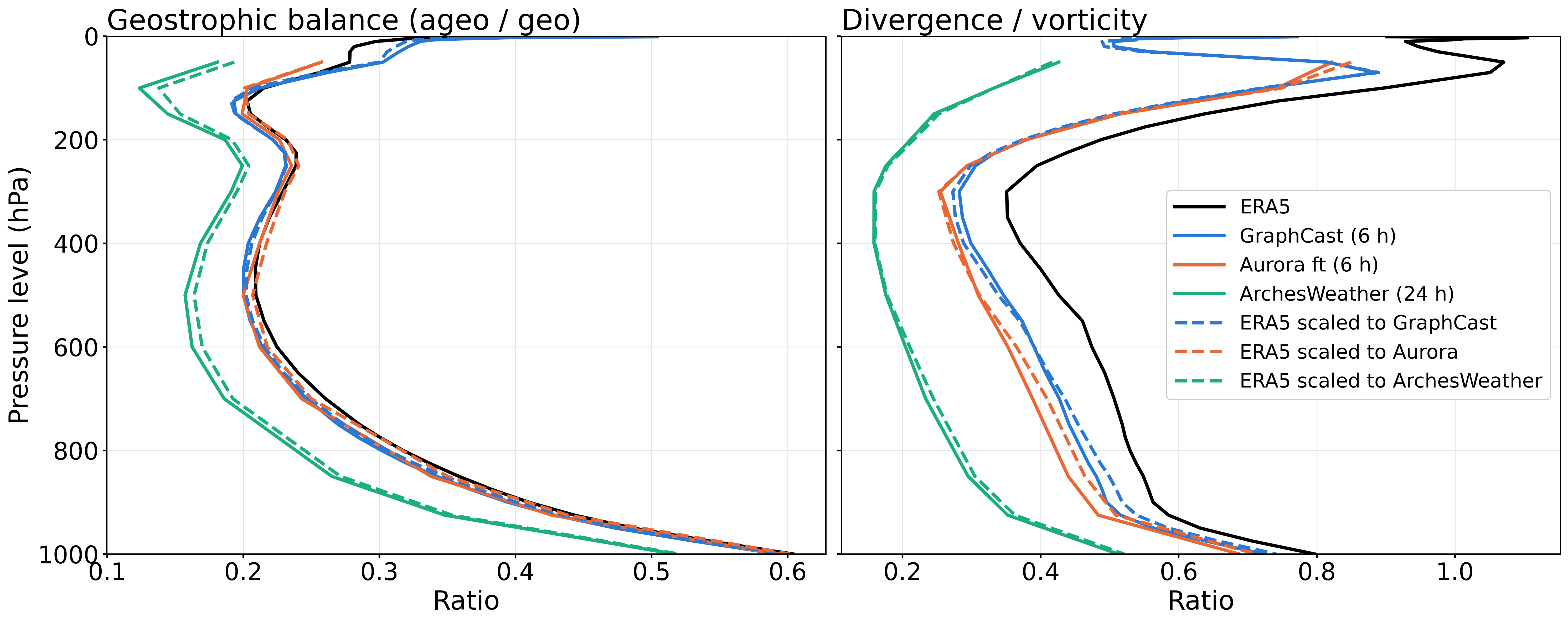}
    \caption{Vertical profiles of physical balances for the full model runs, averaged over the 48 evaluation runs (the 10th of each month in 2020 at 00, 06, 12, and 18\,UTC). Top: ratio of ageostrophic to geostrophic wind, $r_\mathrm{geo}$; bottom: ratio of divergence to vorticity, $r_\mathrm{div}$, for all three models and the ERA5 reference. Dotted lines show ERA5 scaled in spectral space to match each model's spectral energy at each total wavenumber $\ell$.}
    \label{fig:physicalbalances}
\end{figure*}

\begin{figure*}[!t]
    \centering
    \includegraphics[width=1\linewidth]{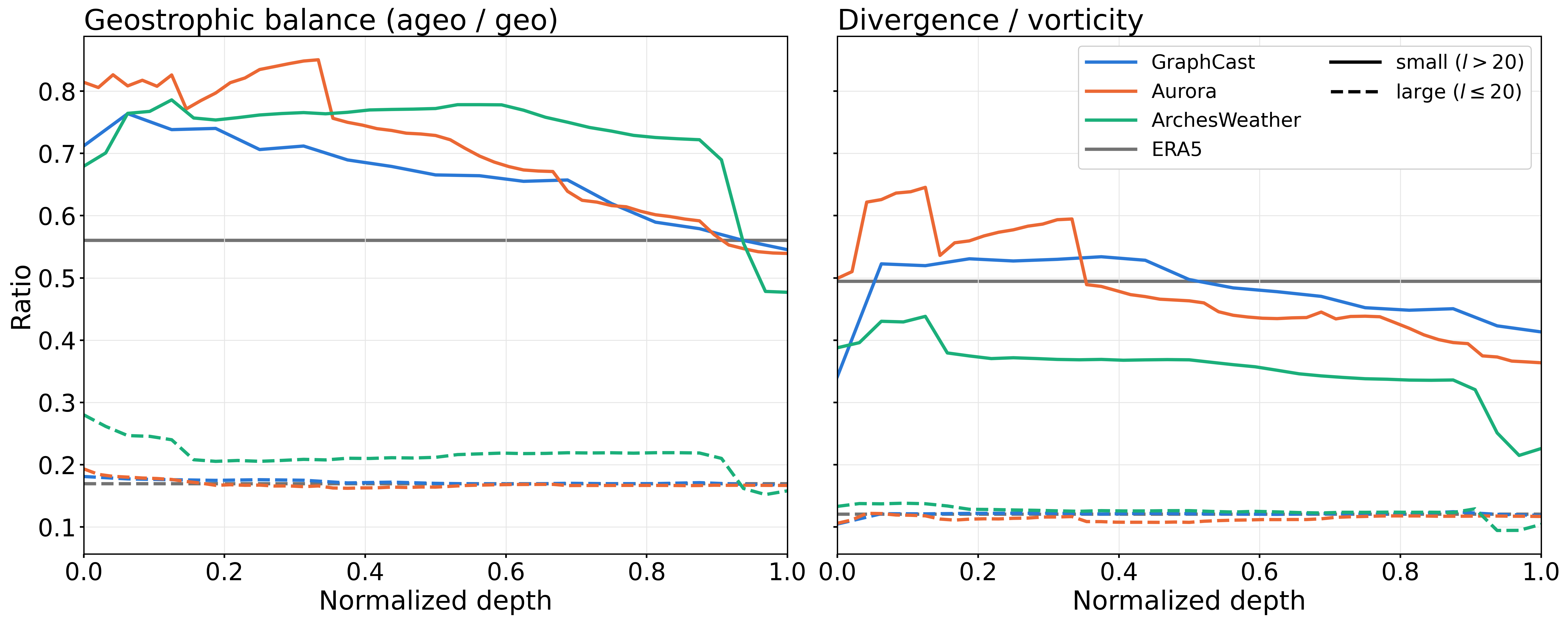}
    \caption{Ratio of ageostrophic to geostrophic wind, $r_\mathrm{geo}$, and ratio of divergence to vorticity, $r_\mathrm{div}$, at 500\,hPa for the intermediate decodings of GraphCast, Aurora, and ArchesWeather as a function of model depth, averaged over the 48 evaluation runs (the 10th of each month in 2020 at 00, 06, 12, and 18\,UTC). Dotted lines show large scales ($\ell \leq 20$), solid lines small scales ($\ell > 20$). For reference, horizontal lines show the corresponding values for the two scale ranges in ERA5.}
    \label{fig:balance_intdec}
\end{figure*}

We now examine the physical consistency of the intermediate decodings by testing for geostrophic balance. Following \citet{bonavita_limitations_2024}, we compute two ratios: the ageostrophic-to-geostrophic wind ratio,
\begin{equation}
    r_{\mathrm{geo}}
    =
    \frac{\langle \lVert \mathbf{u}_{\mathrm{ageo}} \rVert\rangle}
    {\langle\lVert \mathbf{u}_{\mathrm{geo}} \rVert\rangle}
    \qquad \text{for } |\varphi| > 20^\circ,
    \label{eq:geo}
\end{equation}
and the divergence-to-vorticity ratio,
\begin{equation}
    r_{\mathrm{div}}
    =
    \frac{\langle|\delta|\rangle}
    {\langle|\zeta|\rangle}
    \qquad \text{for all latitudes},
    \label{eq:div}
\end{equation}
where $\langle\cdot\rangle$ denotes the latitude-weighted mean over all grid points in the respective domain, $\mathbf{u}$ denotes the horizontal wind and $\delta$ and $\zeta$ denote horizontal divergence and relative vorticity. Both ratios are averaged over the 48 evaluation runs.

As a reference, we first compare these ratios to ERA5, using the full AI model forecasts, including all processor blocks. As ageostrophic wind as well as divergence are concentrated at small scales, such a comparison is only meaningful among data with the same resolution. However, the effective resolution of AI weather models is lower than their output resolution, as they are known to lack variance on small scales. We therefore additionally apply spectral scaling to ERA5 following \citet{Selz2026}, smoothing it to match the effective resolution of each AI model. We then examine these balances for intermediate decoding forecasts to see how the physical balances develop when additional processor blocks are included in the forecast. 

Figure~\ref{fig:physicalbalances} shows vertical profiles of $r_\mathrm{geo}$ and $r_\mathrm{div}$ for the three models alongside ERA5 and spectrally scaled ERA5. Comparing the AI models to the full-resolution ERA5 data shows that on most vertical levels, GraphCast and Aurora are slightly closer to geostrophic balance than ERA5, while ArchesWeather is much more balanced, with ratios about 0.1 lower. This is expected, given the coarser $1.5^\circ$ resolution of ArchesWeather. 
Comparing with the appropriately smoothed ERA5 data removes most of these differences and shows that all three models have a very similar degree of geostrophic balance to that seen in ERA5.

To see where in the processor that balance is created, Figure~\ref{fig:balance_intdec} shows the progression through the layers of the two ratios $r_\mathrm{geo}$ and $r_\mathrm{div}$ for the 500\,hPa level. Geostrophic balance is only expected to be accurate on synoptic and planetary scales. Therefore we show the ratios for two scale ranges separated by total wavenumber 20, corresponding to a half-wavelength of about 1000~km. 

The larger scales show ratios of 0.2 or less right from the earliest processor layers, indicating that geostrophic balance is intrinsic to the forecast, rather than being introduced as a correction later in the processing. A partial exception to this behavior is $r_\mathrm{geo}$ for ArchesWeather, which starts at 0.3, reduces to 0.25 in the middle layers and reaches the final value of 0.2 only in the final stage of processing. In contrast, the mesoscales ($\ell > 20$) become more balanced during processing, with both ratios decreasing by about 30\%. In the case of $r_\mathrm{geo}$, the final values are similar to the ERA5 value, while $r_\mathrm{div}$ is smaller.

As in previous results, the stage structure is evident in Figure~\ref{fig:balance_intdec}, with Aurora showing a sharp reduction in values when entering encoding stages 2 and 3, and the high-magnitude increments of the last stage of Aurora improving balance considerably. These results are consistent with the scale progression documented in the previous section: the large scales are established by the first increments and are balanced from the start, while the small scales are added gradually and become balanced only as they are refined by later processor blocks.

\section{Conclusions and Outlook}

In this work, we propose the method of intermediate decoding to examine how AI weather models construct their forecasts. Applying this method to three different models, we quantified several aspects of how the forecast evolves through the layers of the network and showed that while some aspects are of a more universal character, others depend on model architecture.

The essence of intermediate decoding is to use the model's decoder to transform the latent space variables to physical quantities. This can only succeed if the model possesses representational stability, that is, that the coordinate directions in the latent space have the same meaning through all layers of the network. This is not guaranteed by the model architecture and is indeed found not to hold for the models we examined. However, if the change in representation amounts to a change of basis in the latent space, it can be corrected by training a translator, an affine transformation that maps the latent space variables at a given stage in the processing to the final representation that the decoder expects. This appears to work in practice, with translated latent space vectors showing consistent structure through the model, and a high degree of covariance alignment. It should be noted, however, that covariance alignment is not conclusive evidence of representational stability. It can be dominated by a subset of high-variance directions, while being insensitive to physically important but less salient features. The decrease in covariance similarity for latent states that are separated by more processor blocks may be evidence that the model represents different features at different stages, and that these features cannot be mapped to one another by an affine transformation. This would represent a fundamental limitation of our method of intermediate decoding, so it is encouraging that a high degree of consistency is found in practice.

The most significant outcome of our investigation is that the models we examined showed a number of universal behaviors, despite differences in architecture and number of parameters. First, the encoder and decoder do not form an inverse pair: decoding the initial encoded representation already yields large-scale, physically meaningful increments, demonstrating that the encoded representation is not merely a compressed version of the input but already constitutes a first prediction that can be extracted. The processor then refines this initial prediction, with loss decreasing near-linearly as successive processor blocks are included across model depth, and with a distinct stage structure in which loss reduces most strongly at the beginning of each stage, consistent with iterative refinement, as a new representation is introduced into the forecast. Additionally, the rates at which individual variables are predicted show strong similarities across models, and spatial fields provide examples of this iterative refinement for individual features, such as fronts, across all models. Successive processor blocks further progress from large to small scales, reflecting that atmospheric variance, and consequently loss, is concentrated at large scales. Finally, the physical balances show a similar picture of scale progression: large scales remain physically balanced following the encoder increment and continue to do so with each subsequently added increment, while small scales progressively approach balance with each additional processor block.

In contrast to the universal behaviors described in the previous paragraph, the behaviors that differ between models appear directly tied to specific architectural details. A first example is the abrupt changes in representation at the boundaries of model layers where downsampling or upsampling occurs. These boundaries also lead to changes in the rate of convergence of the forecast to its final value, which is also influenced by skip connections to earlier processor layers. Finally, a more mundane model-dependent behavior is the appearance of noise in patterns that reflect the models' spatial mesh or attention windows. An interesting question is whether these behaviors will diminish with increasing model size or training, analogous to truncation errors in conventional numerical models, allowing convergence to an underlying universal behavior.

Intermediate decoding is a method to interpret the latent state of an AI weather model, specifically to decode the latent representation to physical variables. This provides useful insight into the model behavior, but does not directly relate the evolution of the model state to the learned parameters of the model. The work presented here is only an early step toward improving the interpretability and trustworthiness of AI-based weather forecasting systems.

\acknowledgments
The authors thank the Physics department at LMU for providing computing 
resources and support, and acknowledge fruitful discussions with colleagues. 

\paragraph*{AI use statement.}

In this work, we used generative AI tools for code generation and revising the manuscript. Claude Code (Anthropic) was used to create portions of the analysis and plotting code. Claude Sonnet (Anthropic) was also used to suggest revisions to the manuscript. The research ideas and scientific questions were developed without generative AI assistance. Furthermore, AI was not used to interpret the results. All AI-assisted code and text were reviewed and edited by the authors. We take responsibility for the final content of this work.

\datastatement

The results in this study are based on latent states extracted from all 6-hourly forecasts for the year 2020 and on the intermediate decodings for 48 initialization times (the 10th of each month in 2020 at 00, 06, 12 and 18~UTC) for all three models. These amount to about 17~TB of data and are therefore not published. All of these data can be recreated with the publicly available models on GitHub: \texttt{google-deepmind/graphcast} (commit \texttt{c037656}), \texttt{microsoft/aurora} (release v1.8.0, commit \texttt{fac03c7}) and \texttt{gcouairon/ArchesWeather} (commit \texttt{17e08b1}), using ERA5 data as input. In Aurora and ArchesWeather, latent states can be extracted by using PyTorch hooks, whereas in GraphCast the processor code has to be modified. Intermediate decodings are obtained by passing translated or raw latent states to the decoder.

The code to reproduce the figures will be available at
\url{https://github.com/MatthiasBey/intermediate-decoding/} and is archived at
\url{https://doi.org/10.5281/zenodo.23422846} under the MIT licence. The precomputed diagnostics and
data are available under the CC BY 4.0 licence at
\url{https://doi.org/10.5281/zenodo.23345097},
\url{https://doi.org/10.5281/zenodo.23345099} and
\url{https://doi.org/10.5281/zenodo.23345101}. We provide the precomputed diagnostics for:

\begin{itemize}
    \item the covariance matrices of the latent states for 2020 (Section~\ref{sec:results}\ref{subsection:represstab}),
    \item the mean loss and relative MSE over the 48 forecasts (Section~\ref{sec:results}\ref{subsection:performance}),
    \item the mean power spectra of the increments over the 48 forecasts (Section~\ref{sec:results}\ref{subsection:scaleprogression}),
    \item the mean physical balances of intermediate decodings over the 48 forecasts (Section~\ref{sec:results}\ref{subsec:phybalances}).
\end{itemize}

Additionally, we provide the data for the intermediate decoding forecasts and extracted latent states for the single forecast valid at 12:00~UTC on 10 March 2020 for all three models. A calculation pipeline is provided, which allows recomputing the above mean diagnostics, except for the covariance matrices, for this single forecast.

We further provide the trained translators and their training logs.

ERA5 data are provided by the Copernicus Climate Change Service (C3S).

%






%





\bibliographystyle{ametsocV6}
\bibliography{references}


\end{document}

%% file: Figures/ResNet_diagram.pdf_tex
\begingroup%
  \makeatletter%
  \providecommand\color[2][]{%
    \errmessage{(Inkscape) Color is used for the text in Inkscape, but the package 'color.sty' is not loaded}%
    \renewcommand\color[2][]{}%
  }%
  \providecommand\transparent[1]{%
    \errmessage{(Inkscape) Transparency is used (non-zero) for the text in Inkscape, but the package 'transparent.sty' is not loaded}%
    \renewcommand\transparent[1]{}%
  }%
  \providecommand\rotatebox[2]{#2}%
  \newcommand*\fsize{\dimexpr\f@size pt\relax}%
  \newcommand*\lineheight[1]{\fontsize{\fsize}{#1\fsize}\selectfont}%
  \ifx\svgwidth\undefined%
    \setlength{\unitlength}{595.27559055bp}%
    \ifx\svgscale\undefined%
      \relax%
    \else%
      \setlength{\unitlength}{\unitlength * \real{\svgscale}}%
    \fi%
  \else%
    \setlength{\unitlength}{\svgwidth}%
  \fi%
  \global\let\svgwidth\undefined%
  \global\let\svgscale\undefined%
  \makeatother%
  \begin{picture}(1,1.41428571)%
    \lineheight{1}%
    \setlength\tabcolsep{0pt}%
    \put(0,0){\includegraphics[width=\unitlength,page=1]{ResNet_diagram.pdf}}%
    \put(0.4503059,1.35031949){\color[rgb]{0,0,0}\makebox(0,0)[lt]{\lineheight{1.25}\smash{\begin{tabular}[t]{l}$F_i(\tilde{X}_i)$\end{tabular}}}}%
    \put(0,0){\includegraphics[width=\unitlength,page=2]{ResNet_diagram.pdf}}%
    \put(0.66383267,1.35031949){\color[rgb]{0,0,0}\makebox(0,0)[lt]{\lineheight{1.25}\smash{\begin{tabular}[t]{l}$F_L(\tilde{X}_L)$\end{tabular}}}}%
    \put(0.16032987,1.21413014){\color[rgb]{0,0,0}\makebox(0,0)[lt]{\lineheight{1.25}\smash{\begin{tabular}[t]{l}$\tilde{X}_0$\end{tabular}}}}%
    \put(0.34182531,1.21413014){\color[rgb]{0,0,0}\makebox(0,0)[lt]{\lineheight{1.25}\smash{\begin{tabular}[t]{l}$\tilde{X}_1$\end{tabular}}}}%
    \put(0.55211957,1.21413014){\color[rgb]{0,0,0}\makebox(0,0)[lt]{\lineheight{1.25}\smash{\begin{tabular}[t]{l}$\tilde{X}_i$\end{tabular}}}}%
    \put(0.7699733,1.21413014){\color[rgb]{0,0,0}\makebox(0,0)[lt]{\lineheight{1.25}\smash{\begin{tabular}[t]{l}$\tilde{X}_L$\end{tabular}}}}%
    \put(0.02889666,1.19111423){\color[rgb]{0,0,0}\makebox(0,0)[lt]{\lineheight{1.25}\smash{\begin{tabular}[t]{l}Encoder\end{tabular}}}}%
    \put(0.84555888,1.19111423){\color[rgb]{0,0,0}\makebox(0,0)[lt]{\lineheight{1.25}\smash{\begin{tabular}[t]{l}Decoder\end{tabular}}}}%
    \put(0,0){\includegraphics[width=\unitlength,page=3]{ResNet_diagram.pdf}}%
    \put(0.24049578,1.35031949){\color[rgb]{0,0,0}\makebox(0,0)[lt]{\lineheight{1.25}\smash{\begin{tabular}[t]{l}$F_0(\tilde{X}_0)$\end{tabular}}}}%
    \put(0,0){\includegraphics[width=\unitlength,page=4]{ResNet_diagram.pdf}}%
  \end{picture}%
\endgroup%

%% file: Figures/architecture_diagram.pdf_tex
\begingroup%
  \makeatletter%
  \providecommand\color[2][]{%
    \errmessage{(Inkscape) Color is used for the text in Inkscape, but the package 'color.sty' is not loaded}%
    \renewcommand\color[2][]{}%
  }%
  \providecommand\transparent[1]{%
    \errmessage{(Inkscape) Transparency is used (non-zero) for the text in Inkscape, but the package 'transparent.sty' is not loaded}%
    \renewcommand\transparent[1]{}%
  }%
  \providecommand\rotatebox[2]{#2}%
  \newcommand*\fsize{\dimexpr\f@size pt\relax}%
  \newcommand*\lineheight[1]{\fontsize{\fsize}{#1\fsize}\selectfont}%
  \ifx\svgwidth\undefined%
    \setlength{\unitlength}{595.27559055bp}%
    \ifx\svgscale\undefined%
      \relax%
    \else%
      \setlength{\unitlength}{\unitlength * \real{\svgscale}}%
    \fi%
  \else%
    \setlength{\unitlength}{\svgwidth}%
  \fi%
  \global\let\svgwidth\undefined%
  \global\let\svgscale\undefined%
  \makeatother%
  \begin{picture}(1,1.41428571)%
    \lineheight{1}%
    \setlength\tabcolsep{0pt}%
    \put(0.0345608,1.36226491){\color[rgb]{0,0,0}\makebox(0,0)[lt]{\lineheight{1.25}\smash{\begin{tabular}[t]{l}\large{Graphcast}\end{tabular}}}}%
    \put(0,0){\includegraphics[width=\unitlength,page=1]{architecture_diagram.pdf}}%
    \put(0.13881406,1.298308){\color[rgb]{0,0,0}\makebox(0,0)[lt]{\lineheight{1.25}\smash{\begin{tabular}[t]{l}GNN\end{tabular}}}}%
    \put(0.13648699,1.32979493){\color[rgb]{0,0,0}\makebox(0,0)[lt]{\lineheight{1.25}\smash{\begin{tabular}[t]{l}$\tilde{X}_1 - \tilde{X}_{16}$\end{tabular}}}}%
    \put(0.02893754,1.298308){\color[rgb]{0,0,0}\makebox(0,0)[lt]{\lineheight{1.25}\smash{\begin{tabular}[t]{l}Encoder\end{tabular}}}}%
    \put(0.24798517,1.298308){\color[rgb]{0,0,0}\makebox(0,0)[lt]{\lineheight{1.25}\smash{\begin{tabular}[t]{l}Decoder\end{tabular}}}}%
    \put(0,0){\includegraphics[width=\unitlength,page=2]{architecture_diagram.pdf}}%
    \put(0.13194058,1.26262782){\color[rgb]{0,0,0}\makebox(0,0)[lt]{\lineheight{1.25}\smash{\begin{tabular}[t]{l}$ \mathbb{R}^{40962 \times 512}$\end{tabular}}}}%
    \put(0.03561647,1.04524896){\color[rgb]{0,0,0}\makebox(0,0)[lt]{\lineheight{1.25}\smash{\begin{tabular}[t]{l}\large{ArchesWeather}\end{tabular}}}}%
    \put(0,0){\includegraphics[width=\unitlength,page=3]{architecture_diagram.pdf}}%
    \put(0.50630842,0.95545295){\color[rgb]{0,0,0}\makebox(0,0)[lt]{\lineheight{1.25}\smash{\begin{tabular}[t]{l}conc\end{tabular}}}}%
    \put(0.18154893,1.00736614){\color[rgb]{0,0,0}\makebox(0,0)[lt]{\lineheight{1.25}\smash{\begin{tabular}[t]{l}Skip\end{tabular}}}}%
    \put(0.50535841,1.00882474){\color[rgb]{0,0,0}\makebox(0,0)[lt]{\lineheight{1.25}\smash{\begin{tabular}[t]{l}Skip\end{tabular}}}}%
    \put(0.12023641,0.98547371){\color[rgb]{0,0,0}\makebox(0,0)[lt]{\lineheight{1.25}\smash{\begin{tabular}[t]{l}$\tilde{X}_1 - \tilde{X}_{4}$\end{tabular}}}}%
    \put(0.26852117,0.98541968){\color[rgb]{0,0,0}\makebox(0,0)[lt]{\lineheight{1.25}\smash{\begin{tabular}[t]{l}$\tilde{X}_5 - \tilde{X}_{16}$\end{tabular}}}}%
    \put(0.36505842,0.98483149){\color[rgb]{0,0,0}\makebox(0,0)[lt]{\lineheight{1.25}\smash{\begin{tabular}[t]{l}$\tilde{X}_{17} - \tilde{X}_{28}$\end{tabular}}}}%
    \put(0.57458223,0.9855278){\color[rgb]{0,0,0}\makebox(0,0)[lt]{\lineheight{1.25}\smash{\begin{tabular}[t]{l}$\tilde{X}_{29} - \tilde{X}_{32}$\end{tabular}}}}%
    \put(0.02893754,0.95334451){\color[rgb]{0,0,0}\makebox(0,0)[lt]{\lineheight{1.25}\smash{\begin{tabular}[t]{l}Encoder\end{tabular}}}}%
    \put(0.12401077,0.95532216){\color[rgb]{0,0,0}\makebox(0,0)[lt]{\lineheight{1.25}\smash{\begin{tabular}[t]{l}Layer 1\end{tabular}}}}%
    \put(0.27162334,0.95532216){\color[rgb]{0,0,0}\makebox(0,0)[lt]{\lineheight{1.25}\smash{\begin{tabular}[t]{l}Layer 2\end{tabular}}}}%
    \put(0.36866935,0.95532222){\color[rgb]{0,0,0}\makebox(0,0)[lt]{\lineheight{1.25}\smash{\begin{tabular}[t]{l}Layer 3\end{tabular}}}}%
    \put(0.57811079,0.95532216){\color[rgb]{0,0,0}\makebox(0,0)[lt]{\lineheight{1.25}\smash{\begin{tabular}[t]{l}Layer 4\end{tabular}}}}%
    \put(0.67162984,0.95341739){\color[rgb]{0,0,0}\makebox(0,0)[lt]{\lineheight{1.25}\smash{\begin{tabular}[t]{l}Decoder\end{tabular}}}}%
    \put(0.10672877,0.91751958){\color[rgb]{0,0,0}\makebox(0,0)[lt]{\lineheight{1.25}\smash{\begin{tabular}[t]{l}$ \mathbb{R}^{8 \times 60 \times 120 \times 192}$\end{tabular}}}}%
    \put(0.25470951,0.91751943){\color[rgb]{0,0,0}\makebox(0,0)[lt]{\lineheight{1.25}\smash{\begin{tabular}[t]{l}$ \mathbb{R}^{8 \times 30 \times 60 \times 384}$\end{tabular}}}}%
    \put(0.35897672,0.91751943){\color[rgb]{0,0,0}\makebox(0,0)[lt]{\lineheight{1.25}\smash{\begin{tabular}[t]{l}$ \mathbb{R}^{8 \times 30 \times 60 \times 384}$\end{tabular}}}}%
    \put(0.55807255,0.91751932){\color[rgb]{0,0,0}\makebox(0,0)[lt]{\lineheight{1.25}\smash{\begin{tabular}[t]{l}$ \mathbb{R}^{8 \times 60 \times 120 \times 384}$\end{tabular}}}}%
    \put(0,0){\includegraphics[width=\unitlength,page=4]{architecture_diagram.pdf}}%
    \put(0.76643496,1.28364526){\color[rgb]{0,0,0}\makebox(0,0)[lt]{\lineheight{1.25}\smash{\begin{tabular}[t]{l}upscale\end{tabular}}}}%
    \put(0,0){\includegraphics[width=\unitlength,page=5]{architecture_diagram.pdf}}%
    \put(0.76670752,1.33798862){\color[rgb]{0,0,0}\makebox(0,0)[lt]{\lineheight{1.25}\smash{\begin{tabular}[t]{l}downscale\end{tabular}}}}%
    \put(0.88817928,1.29090013){\color[rgb]{0,0,0}\makebox(0,0)[lt]{\lineheight{1.25}\smash{\begin{tabular}[t]{l}concatenate\\channels\end{tabular}}}}%
    \put(0,0){\includegraphics[width=\unitlength,page=6]{architecture_diagram.pdf}}%
    \put(0.84658789,1.27957194){\color[rgb]{0,0,0}\makebox(0,0)[lt]{\lineheight{1.25}\smash{\begin{tabular}[t]{l}conc\end{tabular}}}}%
    \put(0.88528883,1.34640278){\color[rgb]{0,0,0}\makebox(0,0)[lt]{\lineheight{1.25}\smash{\begin{tabular}[t]{l}add\\channels\end{tabular}}}}%
    \put(0,0){\includegraphics[width=\unitlength,page=7]{architecture_diagram.pdf}}%
    \put(0.66898048,1.34635759){\color[rgb]{0,0,0}\makebox(0,0)[lt]{\lineheight{1.25}\smash{\begin{tabular}[t]{l}Encoder/\\Decoder\end{tabular}}}}%
    \put(0,0){\includegraphics[width=\unitlength,page=8]{architecture_diagram.pdf}}%
    \put(0.67622746,1.2930991){\color[rgb]{0,0,0}\makebox(0,0)[lt]{\lineheight{1.25}\smash{\begin{tabular}[t]{l}Processor\\layer\end{tabular}}}}%
    \put(0.03561647,1.22350424){\color[rgb]{0,0,0}\makebox(0,0)[lt]{\lineheight{1.25}\smash{\begin{tabular}[t]{l}\large{Aurora}\end{tabular}}}}%
    \put(0,0){\includegraphics[width=\unitlength,page=9]{architecture_diagram.pdf}}%
    \put(0.17056145,1.18537332){\color[rgb]{0,0,0}\makebox(0,0)[lt]{\lineheight{1.25}\smash{\begin{tabular}[t]{l}Skip 1\end{tabular}}}}%
    \put(0.84800727,1.18292734){\color[rgb]{0,0,0}\makebox(0,0)[lt]{\lineheight{1.25}\smash{\begin{tabular}[t]{l}Skip 1\end{tabular}}}}%
    \put(0.85714286,1.12837466){\color[rgb]{0,0,0}\makebox(0,0)[lt]{\lineheight{1.25}\smash{\begin{tabular}[t]{l}conc\end{tabular}}}}%
    \put(0.11127386,1.15775721){\color[rgb]{0,0,0}\makebox(0,0)[lt]{\lineheight{1.25}\smash{\begin{tabular}[t]{l}$\tilde{X}_1 - \tilde{X}_{6}$\end{tabular}}}}%
    \put(0.25581238,1.15711136){\color[rgb]{0,0,0}\makebox(0,0)[lt]{\lineheight{1.25}\smash{\begin{tabular}[t]{l}$\tilde{X}_7 - \tilde{X}_{16}$\end{tabular}}}}%
    \put(0.38212974,1.15702962){\color[rgb]{0,0,0}\makebox(0,0)[lt]{\lineheight{1.25}\smash{\begin{tabular}[t]{l}$\tilde{X}_{17} - \tilde{X}_{24}$\end{tabular}}}}%
    \put(0.46807462,1.15711278){\color[rgb]{0,0,0}\makebox(0,0)[lt]{\lineheight{1.25}\smash{\begin{tabular}[t]{l}$\tilde{X}_{25} - \tilde{X}_{32}$\end{tabular}}}}%
    \put(0.76890876,1.15690526){\color[rgb]{0,0,0}\makebox(0,0)[lt]{\lineheight{1.25}\smash{\begin{tabular}[t]{l}$\tilde{X}_{43} - \tilde{X}_{48}$\end{tabular}}}}%
    \put(0.02898691,1.124446){\color[rgb]{0,0,0}\makebox(0,0)[lt]{\lineheight{1.25}\smash{\begin{tabular}[t]{l}Encoder\end{tabular}}}}%
    \put(0.91332565,1.12627029){\color[rgb]{0,0,0}\makebox(0,0)[lt]{\lineheight{1.25}\smash{\begin{tabular}[t]{l}Decoder\end{tabular}}}}%
    \put(0.11196711,1.12494598){\color[rgb]{0,0,0}\makebox(0,0)[lt]{\lineheight{1.25}\smash{\begin{tabular}[t]{l}Enc 1\end{tabular}}}}%
    \put(0.25986395,1.12494598){\color[rgb]{0,0,0}\makebox(0,0)[lt]{\lineheight{1.25}\smash{\begin{tabular}[t]{l}Enc 2\end{tabular}}}}%
    \put(0.38795151,1.12494598){\color[rgb]{0,0,0}\makebox(0,0)[lt]{\lineheight{1.25}\smash{\begin{tabular}[t]{l}Enc 3\end{tabular}}}}%
    \put(0.47359408,1.12459819){\color[rgb]{0,0,0}\makebox(0,0)[lt]{\lineheight{1.25}\smash{\begin{tabular}[t]{l}Dec 1\end{tabular}}}}%
    \put(0.77507557,1.12494598){\color[rgb]{0,0,0}\makebox(0,0)[lt]{\lineheight{1.25}\smash{\begin{tabular}[t]{l}Dec 3\end{tabular}}}}%
    \put(0,0){\includegraphics[width=\unitlength,page=10]{architecture_diagram.pdf}}%
    \put(0.09846735,1.08959737){\color[rgb]{0,0,0}\makebox(0,0)[lt]{\lineheight{1.25}\smash{\begin{tabular}[t]{l}$ \mathbb{R}^{4 \times 180 \times 360 \times 512}$\end{tabular}}}}%
    \put(0.2327889,1.08959737){\color[rgb]{0,0,0}\makebox(0,0)[lt]{\lineheight{1.25}\smash{\begin{tabular}[t]{l}$ \mathbb{R}^{4 \times 90 \times 180 \times 1024}$\end{tabular}}}}%
    \put(0.36234375,1.08959738){\color[rgb]{0,0,0}\makebox(0,0)[lt]{\lineheight{1.25}\smash{\begin{tabular}[t]{l}$ \mathbb{R}^{4 \times 45 \times 90 \times 2048}$\end{tabular}}}}%
    \put(0.74962622,1.08959737){\color[rgb]{0,0,0}\makebox(0,0)[lt]{\lineheight{1.25}\smash{\begin{tabular}[t]{l}$ \mathbb{R}^{4 \times 180 \times 360 \times 512}$\end{tabular}}}}%
    \put(0.46339133,1.08959738){\color[rgb]{0,0,0}\makebox(0,0)[lt]{\lineheight{1.25}\smash{\begin{tabular}[t]{l}$ \mathbb{R}^{4 \times 45 \times 90 \times 2048}$\end{tabular}}}}%
    \put(0.57604526,1.08959737){\color[rgb]{0,0,0}\makebox(0,0)[lt]{\lineheight{1.25}\smash{\begin{tabular}[t]{l}$ \mathbb{R}^{4 \times 90 \times 180 \times 1024}$\end{tabular}}}}%
    \put(0,0){\includegraphics[width=\unitlength,page=11]{architecture_diagram.pdf}}%
    \put(0.59466329,1.15696735){\color[rgb]{0,0,0}\makebox(0,0)[lt]{\lineheight{1.25}\smash{\begin{tabular}[t]{l}$\tilde{X}_{33} - \tilde{X}_{42}$\end{tabular}}}}%
    \put(0.59984834,1.12494598){\color[rgb]{0,0,0}\makebox(0,0)[lt]{\lineheight{1.25}\smash{\begin{tabular}[t]{l}Dec 2\end{tabular}}}}%
    \put(0,0){\includegraphics[width=\unitlength,page=12]{architecture_diagram.pdf}}%
    \put(0.70499741,1.1834488){\color[rgb]{0,0,0}\makebox(0,0)[lt]{\lineheight{1.25}\smash{\begin{tabular}[t]{l}Skip 1\\\end{tabular}}}}%
    \put(0,0){\includegraphics[width=\unitlength,page=13]{architecture_diagram.pdf}}%
  \end{picture}%
\endgroup%